\documentclass[a4paper,11pt]{article}

\usepackage[utf8]{inputenc}
\usepackage[english]{babel}
\usepackage[myheadings]{fullpage}
\usepackage[T1]{fontenc}
\usepackage{subfig}
\usepackage{fancyhdr}
\usepackage{graphicx}
\usepackage{sectsty}
\usepackage{url}
\usepackage[usenames, dvipsnames, table, xcdraw]{xcolor}
\usepackage{amsmath}
\usepackage{amssymb}
\usepackage{calrsfs}
\usepackage{bbold}
\usepackage{dsfont}
\usepackage{esint}
\usepackage{indentfirst}
\usepackage{xfrac}
\usepackage{bigints}
\usepackage{slashed}
\usepackage{breqn}
\usepackage{autobreak}
\usepackage{setspace}
\usepackage[colorlinks]{hyperref}
\usepackage{microtype}
\usepackage{quoting}
\usepackage{float}
\usepackage{array}
\usepackage[export]{adjustbox}
\usepackage[section]{placeins}
\usepackage[capitalise]{cleveref}
\usepackage{enumitem}
\usepackage{scalefnt}
\usepackage{booktabs}
\usepackage{caption}
\usepackage{makecell}
\usepackage{authblk}
\usepackage{cases}
\usepackage{multirow}

\usepackage[normalem]{ulem}

\usepackage{cite}

\usepackage{xspace}

\usepackage{colortbl}
\definecolor{myblue}{rgb}{0.8,0.85,1}
\definecolor{light-gray}{gray}{0.95}

\hypersetup{colorlinks=true,
citecolor=ForestGreen,
anchorcolor=CornflowerBlue,
linkcolor=NavyBlue,
urlcolor=Mahogany,
linkbordercolor={1 1 0}}

\DeclareMathAlphabet{\mathcal}{OMS}{cmsy}{m}{n}

\makeatletter
\newcommand{\thickhline}{\noalign {\ifnum 0=`}\fi \hrule height 1pt
\futurelet \reserved@a \@xhline}
\newcolumntype{"}{@{\hskip\tabcolsep\vrule width 1pt\hskip\tabcolsep}}
\makeatother

\allowdisplaybreaks

\def\beq {\begin{equation}}
\def\eeq {\end{equation}}
\def\bea {\begin{eqnarray}}
\def\eea {\end{eqnarray}}

\AtBeginDocument{}

\title{\LARGE {\bf \sffamily \boldmath Model-independent extrapolation of MUonE data with Pad\'e and D-Log approximants}}

\author[a]{Diogo Boito}
\author[a,b]{Cristiane Y. London}
\author[b]{Pere Masjuan}
\author[b]{Camilo Rojas}

\affil[a]{\it Instituto de F\'isica de S\~ao Carlos, Universidade de S\~ao Paulo, CP 369, 13560-970, S\~ao Carlos, SP, Brazil\vspace{0.2cm}}

\affil[b]{\it Grup de F\'isica Te\`orica, Departament de F\'isica, Universitat Aut\`onoma de Barcelona, and Institut de F\'isica d'Altes Energies (IFAE), The Barcelona Institute of Science and Technology (BIST), Campus UAB, E-08193 Bellaterra (Barcelona), Spain\vspace{0.3cm}}

\date{}

\begin{document}
\begin{flushright}
{\small \today}
\end{flushright}

\vspace*{-0.7cm}
\begingroup
\let\newpage\relax
\maketitle
\endgroup
\date{}

\vspace*{-1.0cm}
\begin{abstract}
\noindent
The MUonE experiment is designed to extract the hadronic contribution to the electromagnetic coupling in the space-like region, $\Delta \alpha_{\rm had}(t)$, from elastic $e\mu$ scattering.
The leading order hadronic vacuum polarization contribution to the muon $g-2$,
$a_\mu^{\mathrm{HVP, \,LO}}$, can then be obtained from a weighted integral over $\Delta \alpha_{\rm had}(t)$. This, however, requires knowledge of $\Delta \alpha_{\rm had}(t)$ in the whole domain of integration, which cannot be achieved by experiment. In this work, we propose to use Pad\'e and D-Log Pad\'e approximants as a systematic and model-independent method to fit and reliably extrapolate the future MUonE experimental data, extracting $a_\mu^{\mathrm{HVP,\,LO}}$ with a conservative but competitive uncertainty, using no or very limited external information. The method relies on fundamental analytic properties of the two-point correlator underlying $a_\mu^{\mathrm{HVP,\,LO}}$ and provides
lower and upper bounds for the result for $a_\mu^{\mathrm{HVP,\,LO}}$. We demonstrate the reliability of the method using toy data sets generated from a model for $\Delta \alpha_{\rm had}(t)$ reflecting the expected statistics of the MUonE experiment.

\end{abstract}

\thispagestyle{empty}

\vspace*{0.0cm}
\newpage
\tableofcontents

\setcounter{page}{1}

\vspace{1cm}

\section{Introduction}
The recent measurements of the anomalous magnetic moment of the muon, $a_\mu=(g-2)/2$, by the FNAL E989 experiment at Fermilab, in 2021 and 2023~\cite{Muong-2:2021ojo,Muong-2:2023cdq}, are in good agreement with the previous experimental result from the Brookhaven National Lab BNL E821 experiment of 2006~\cite{Muong-2:2006rrc}.
The combination of the results leads to an experimental determination of $a_\mu$ with an impressive uncertainty of only 0.19 ppm.
As is well known, the 2020 $g-2$ Theory Initiative White Paper~\cite{Aoyama:2020ynm} recommended result for $a_\mu$ in the Standard Model (based on the results of Refs.~\cite{Aoyama:2012wk,Aoyama:2019ryr,Czarnecki:2002nt,Gnendiger:2013pva,Davier:2017zfy,Keshavarzi:2018mgv,Davier:2019can,Keshavarzi:2019abf,Colangelo:2018mtw,Hoferichter:2019mqg,Hoid:2020xjs,Kurz:2014wya,Melnikov:2003xd,Masjuan:2017tvw,Colangelo:2017qdm,Colangelo:2017fiz,Hoferichter:2018dmo,Hoferichter:2018kwz,Gerardin:2019vio,Bijnens:2019ghy,Colangelo:2019lpu,Colangelo:2019uex,Colangelo:2014qya,Blum:2019ugy}) is 5.1$\sigma$ lower than the new, combined, experimental number --- a tension that has attracted enormous attention in the past few years. This result relies on the dispersive description of the Hadronic Vacuum Polarization (HVP) contribution to $a_\mu$, $a_\mu^{\rm HVP}$. If one employs instead the recent lattice QCD results for $a_\mu^{\rm HVP}$ obtained by the BMW Collaboration~\cite{Borsanyi:2020mff}, the discrepancy between theory and experiment would be reduced to 2.0$\sigma$.

Understanding the origin of the tension between the dispersive-based result and the lattice-based determination of $a_\mu^{\rm HVP}$ is of crucial importance. The detailed comparison is not completely straightforward~\cite{Hansen:2019idp,Bailas:2020qmv,Boito:2022njs} since in lattice QCD one has access to the Euclidean HVP, while the dispersive approach relies on data for $e^+e^-\to ({\rm hadrons})$ in the whole Minkowski domain. Independent information from other related processes may be crucial to fully resolve persistent discrepancies. A prominent example is the use of $\tau$ decay data, which requires a non-trivial treatment of isospin corrections~\cite{Miranda:2020wdg,Masjuan:2023qsp}. In this context, the recently proposed MUonE experiment~\cite{CarloniCalame:2015obs,Abbiendi:2016xup,Abbiendi:2677471} would also be very welcome. The proposal is to extract the HVP in the Euclidean domain, directly from data, from the measurement of the elastic $e\mu$ cross-section using the $150$-GeV muon beam from CERN's M2 beamline scattered off atomic electrons of a low-$Z$ target. The experiment could yield competitive results after three years of data taking and would be able to cover approximately 86\%~\cite{Abbiendi:2022oks,Ignatov:2023wma} of the integration interval required for the computation of $a_\mu^{\rm HVP}$ at leading order, $a_\mu^{\rm HVP,\,LO}$.

An important question is how to treat the remaining 14\% of the $a_\mu^{\rm HVP,\,LO}$ integral not directly accessible to the MUonE experiment. In principle, one could simply resort to external information and use the dispersive approach, perturbative QCD, and/or lattice QCD results.
Another option, arguably more interesting, is to extract $a_\mu^{\rm HVP,\,LO}$ exclusively from the MUonE data, which requires some form of extrapolation of the experimental results beyond the kinematically accessible region. This problem is, however, non-trivial since the experiment would have access to a narrow window in the Euclidean $t$ variable, between
$-0.153~{\rm GeV}^2\leq t \leq -0.001~{\rm GeV}^2$. In Ref.~\cite{Abbiendi:2022oks}, a model inspired by one-loop QED is put forward as a fitting function to fit and extrapolate the MUonE data. A disadvantage of this approach is a potential model dependency that could bias the final results. An alternate strategy suggested in Ref.~\cite{Ignatov:2023wma} relies on the extraction of derivatives of the hadronic contribution to the
running of the electromagnetic coupling, supplemented with information from perturbative QCD and $R(s)$ data. Another recent proposal, closer in spirit to our work, consists of using transfer theorems to build so-called ``reconstruction approximants'' which allow for a partial reconstruction of the HVP function to compute $a_\mu^{\rm HVP}$~\cite{Greynat:2022geu}.
Here, we propose the use of Pad\'e approximants (PAs), and a variant of this method, as a systematic, simple, and model-independent way of fitting and extrapolating the MUonE results to compute $a_\mu^{\rm HVP,\,LO}$ from MUonE data.

The extrapolation of the HVP results in the Euclidean domain using PAs has been explored previously in the context of heavy-quark physics~\cite{Masjuan:2009wy} and lattice-QCD results~\cite{Golterman:2014ksa}. The use of PAs in this problem is predicated on the fact that the HVP is a Stieltjes function~\cite{Masjuan:2009wy}. In this case, convergence theorems for sequences of PAs apply and inequalities guarantee that certain PAs' sequences approach the function ``from above'' while others do so ``from below''~\cite{Baker1975essentials,Baker1996pade,Masjuan:2009wy}, providing a systematic way to bound the value of the function of interest. The same theorems play a crucial role in our work.

By construction, the usual PAs are not able to explore the Minkowski region coming from the Euclidean domain as they contain {\it only} poles and zeros, and the branch cuts can be, at best, emulated by the accumulation of singularities~\cite{Baker1996pade, MasjuanQueralt:2010hav, Costin:2021bay}. Therefore, we propose to accompany the PAs study with the use of a variant of the method called D-Log Pad\'e approximants, D-Logs \cite{Boito:2022njs,Boito:2021scm,Boito:2018rwt}, an extension of PAs which contain, by construction, not only poles and zeros but also branch cuts. In some cases, a systematic way to bound the value of the function is also provided pointing towards a similar convergence theorem.
This can pave the way for future explorations where one could have a glimpse of the Minkowski region from fits in the Euclidean region in a model-independent and systematic way.

To assess the reliability and viability of our proposal we adopt a simple but sufficiently realistic model for the HVP function introduced in Ref.~\cite{Greynat:2022geu}. We then build the approximants first to the exact Taylor expansion of the model function and later to pseudo-data generated from the model. In this first step, we do not include uncertainties, as a proof of concept. We then generate realistic pseudo-data following the expected uncertainties and kinematic range accessible to the MUonE experiment~\cite{CarloniCalame:2015obs,Abbiendi:2016xup,Abbiendi:2677471,Ballerini:2019zkk,Abbiendi:2019qtw,Abbiendi:2021xsh,Abbiendi:2022oks}.
With these data sets, we perform a systematic study of the use of PAs and D-Logs, as a way to fit and extrapolate MUonE data.

As we already mentioned, obtaining $a_\mu^{\rm HVP}$ solely from MUonE data is a non-trivial problem. Therefore, we perform a systematic study where we enlarge in each step the window in which we rely on extrapolated results.
In our systematic investigation, we show that there is a trade-off in precision. It is possible to perform a reliable, robust, and model-independent extraction of $a_\mu^{\rm HVP,\,LO}$ using approximants solely from MUonE data, but with a somewhat larger error. An advantage of the method is that both PAs and D-Logs allow for a reliable estimate of the systematic error. If the window in which one uses the extrapolated results is reduced, the error diminishes, as could be expected.

This paper is organized as follows. In Sec.~\ref{sec:theoryHVP}, we introduce the basic elements to describe the HVP contribution to the $(g-2)_{\mu}$ related to the running of the electromagnetic coupling constant. In Sec.~\ref{sec:theorypas}, we review the aspects of Pad\'e Theory that are the foundations of our work, in particular, Stieltjes functions and the convergence theorems of PAs and D-Logs applied to them. In Sec.~\ref{sec:model}, we present the model of Ref.~\cite{Greynat:2022geu} that we use to generate our toy data sets for the Euclidean HVP, while an example of the power of the convergence theorems is presented in Sec.~\ref{sec:pastaylor}, where PAs and D-Logs are built from the exactly known Taylor series given by
the model of Sec.~\ref{sec:model}. In Sec.~\ref{sec:pasnoerror}, we employ our method in the idealized scenario where the data points have zero error. Sec.~\ref{sec:results} illustrates the application of our method to realistic data sets, following the expectations of the MUonE experiment.
Our conclusions are given in Sec.~\ref{sec:conclusion}.
Technical details about the fitting functions are relegated to Appendix~\ref{app:pasfunctions}.

\section{Theoretical framework}\label{sec:theoryHVP}

The Standard Model computation of $a_\mu$ can be divided into four different contributions, namely, from Quantum Electrodynamics (QED), electroweak effects, HVP, and hadronic light-by-light scattering. The dominant uncertainty arises from the HVP contribution~\cite{Aoyama:2020ynm}, more specifically from its leading order contribution, $a_\mu^{\mathrm{HVP, \,LO}}$. In the computation of $a_\mu^{\mathrm{HVP}}$ the main object is the polarization function associated with the electromagnetic current two-point correlator, $\Pi(q^2)$, defined as
\beq
(q_\mu q_\nu -q^2g_{\mu\nu})\Pi(q^2) = i \int d^4x\, \langle 0 | T(j^{\rm EM}_\mu (x) j^{\rm EM}_\nu (0) | 0 \rangle ,
\eeq
where the electromagnetic current is
\beq
j_\mu^{\rm EM}= \frac{2}{3}\bar u \gamma_\mu u -\frac{1}{3} \bar d \gamma_\mu d -\frac{1}{3}\bar s \gamma_\mu s+ \frac{2}{3}\bar c \gamma_\mu c +\cdots .
\eeq
We define $\bar\Pi(q^2)=\Pi(q^2)-\Pi(0)$, and the function $\bar \Pi(q^2)$ obeys the usual once-subtracted dispersion relation
\begin{equation}
\bar{\Pi}(q^2) = \frac{q^2}{\pi} \int_{m_\pi^2}^\infty \mathrm{d}s \dfrac{\operatorname{Im}\Pi(s)}{s(s - q^2 + i \epsilon)} .
\label{eq:dispersiverelation}
\end{equation}

In the dispersive approach, $a_\mu^{\mathrm{HVP, \,LO}}$ is obtained from the inclusive hadronic electroproduction cross-section defined, with $s=q^2$, as
\begin{equation}
R(s) = \left(\dfrac{3s}{4\pi\alpha^2} \right)\,\, \sigma_{e^+e^- \to \mathrm{hadrons}}(s) = 12\pi\, {\rm Im} \Pi(s),
\end{equation}
through the following weighted integral
\begin{equation}
a_\mu^{\mathrm{HVP, \,LO}} = \dfrac{\alpha^2}{3\pi^2} \int_{m_\pi^2}^\infty \dfrac{\mathrm{d}s}{s} \,\, K(s) R(s),
\label{eq:amurs}
\end{equation}
where $\alpha$ is the electromagnetic fine-structure constant and $K(s)$ is the QED kernel function~\cite{Bouchiat:1961lbg,Brodsky:1967sr,Lautrup:1968tdb}
\begin{equation}
K(s) = \int_0^1 \dfrac{x^2 (1-x)}{x^2 + (1-x)\frac{s}{m_\mu^2}} \,\,\, \mathrm{d}x \, .
\end{equation}
The analytical result for $K(s)$ is given explicitly in Ref.~\cite{Aoyama:2020ynm}.

An alternate representation for $a_\mu^{\rm HVP, \, LO}$ in terms of the correlator in the Euclidean, $\Pi(Q^2)$ with $Q^2=-q^2>0$, can be obtained interchanging the order of the integrals in $s$ and $x$ in Eq.~(\ref{eq:amurs})~\cite{Lautrup:1971jf}. Using the analytical properties of $\Pi(q^2)$ one can then write
\begin{equation}
a_\mu^{\mathrm{HVP, \,LO}} = \dfrac{\alpha^2}{\pi} \int_0^1 \mathrm{d}x \,\,\, (1-x) \, \Delta\alpha_{\mathrm{had}}[t(x)],
\label{eq:amu}
\end{equation}
where, following the notation employed by Bernecker and Meyer in Ref.~\cite{Bernecker:2011gh}, we defined
\beq\label{eq:Deltaalpha}
\Delta\alpha_{\mathrm{had}}(t) = - 4 \pi \operatorname{Re}[\bar{\Pi}_{\mathrm{had}}(t)]
\eeq
as the hadronic contribution to the running of the electromagnetic coupling $\alpha$ and $t$ is the space-like variable given by
\begin{equation}
t = - \dfrac{x^2 \, m_\mu^2}{1-x}.
\label{eq:ttox}
\end{equation}

The MUonE experiment is designed to extract $\Delta\alpha_{\mathrm{had}}(t)$ from $e\mu$ scattering data using 150-GeV muons scattered off atomic electrons. This allows, in principle, for a completely
independent determination of $a_\mu^{\rm HVP,\,LO}$.
However, the experiment would be restricted approximately to the window $x \in [0.2, 0.93]$~\cite{CarloniCalame:2015obs,Abbiendi:2016xup,Abbiendi:2677471,Ballerini:2019zkk,Abbiendi:2019qtw,Abbiendi:2021xsh,Abbiendi:2022oks}, which corresponds to $-0.15~{\rm GeV}^2\lesssim t \lesssim -0.001~{\rm GeV}^2$.
To obtain $a_\mu^{\rm HVP,\,LO}$ from MUonE data without requiring external information, it is imperative to have a reliable method to extrapolate the MUonE data way outside the experimentally accessible window. With this purpose in mind, we will use a set of approximants as fitting functions to toy data sets for $\Delta\alpha_{\mathrm{had}}(t)$ simulating the expected results of the MUonE experiment.

Finally, uncertainty assessment is crucial and we want to keep track of the goodness of our extrapolation beyond the fit region. The extrapolation for $0\le x<0.2$ is safe, since it involves a small interval in $t$, not too far from the origin, namely $0\le t\lesssim0.001$~GeV$^2$. The extrapolation for $0.93<x<1$, on the other hand, corresponding to $0.15~{\rm GeV}^2\lesssim t < \infty$, is non-trivial and, in order to assess it carefully, we define $a_\mu^{\mathrm{HVP, \,LO}}(x_{\rm max})$ as the partial contribution from $x=0$ up to $x=x_{\rm max}$ to $a_\mu^{\rm HVP,\,LO}$, expressed as
\begin{equation}
a_\mu^{\mathrm{HVP, \,LO}}(x_{\rm max}) = \dfrac{\alpha^2}{\pi} \int_0^{x_{\rm max}} \mathrm{d}x \,\,\, (1-x) \, \Delta\alpha_{\mathrm{had}}[t(x)].
\label{eq:amuxmax}
\end{equation}

\section{Pad\'e and D-Log Pad\'e approximants}\label{sec:theorypas}

In this section, we give an overview of Pad\'e Theory including Pad\'e approximants and their variant, D-Log Padé approximants, and a discussion of the advantages and disadvantages of each of them in the determination of $a_\mu^{\rm HVP,\,LO}$.

A PA $P_M^N(z)$ is a rational function given by
\begin{equation}
\label{eq:pa}
P_M^N(z) = \dfrac{Q_N(z)}{R_M(z)} = \dfrac{q_0 + q_1 \,z + \cdots + q_N \, z^N}{1 + r_1 \,z + \cdots + r_M \, z^M},
\end{equation}
where we adopted $r_0 = 1$. The standard technique to construct PAs to a function is by matching the first $N+M+1$ terms of the Taylor series expansion of Eq.~(\ref{eq:pa}) to that of the original function order by order, thereby fixing the coefficients of the PA unambiguously~\cite{Baker1975essentials,Baker1996pade,MasjuanQueralt:2010hav}.

A D-Log Pad\'e approximant, to which we refer simply as D-Log, in turn, is a variant that is very useful for functions with branch points or poles with higher multiplicity\cite{Baker1996pade,Boito:2018rwt,Boito:2021scm}. Let us consider, for example, the following function
\beq\label{eq:StieltjesType}
f(z) = A(z) \, \dfrac{1}{(\mu - z)^\gamma} + B(z) ,
\eeq
where $A(z)$ and $B(z)$ are functions with little structure and analytic at $z=\mu$. We are primarily interested in the case in which $f(z)$ has a branch point at $z=\mu$ and, accordingly, $\gamma$ is not necessarily an integer number. We can define now a new function $F(z)$ which, near $z=\mu$, behaves as\cite{Baker1996pade}
\beq
F(z) = \dfrac{\mathrm{d}}{\mathrm{d} z} \ln{f(z)} \approx \dfrac{\gamma}{(\mu - z)} . \label{eq:Fdlog}
\eeq
Even though $\gamma$ does not have to be an integer, $F(z)$ has a simple pole and its residue is the exponent of the cut of $f(z)$. Thus, with $\bar P_M^N(z)$ being the PA constructed to $F(z)$ defined above, the $\mathrm{Dlog}_M^N(z)$, or simply $D_M^N(z)$, of $f(z)$ is given by the expression below\cite{Baker1996pade,Boito:2018rwt,Boito:2021scm}
\beq
\mathrm{Dlog}_M^N(z) \equiv D_M^N(z)= f(0) \exp{\left[\int \mathrm{d}z \, \bar{P}_M^N(z) \right]} . \label{eq:dlogpa}
\eeq
Because of the derivative in \cref{eq:Fdlog}, the $f(0)$ term is lost and must be reintroduced to correctly normalize the D-Log. The $D_M^N$ then reproduces exactly the first $M+N+2$ coefficients of $f(z)$ (one order more than the usual $P_M^N$) and can be used to predict the $(M+N+3)$-th coefficient and higher.

In principle, this type of approximant offers a way to determine the branch point and the exponent of the cut of the original function $f(z)$ from the study of the PA to $F(z)$ around its pole. Since no assumption about $\mu$ or $\gamma$ is made, their estimates are exclusively obtained from the series coefficients.

With these methods, the convergence of sequences of both PAs and D-Logs to the original function is guaranteed for specific types of functions, for example, Stieltjes functions. A Stieltjes function is a function that can be represented by a Stieltjes integral
\begin{equation}
f(z) = \int_0^\infty \dfrac{\mathrm{d}\phi(u)}{1+zu} ,
\end{equation}
where $\phi(u)$ is a bounded non-decreasing function on the interval $0 \leq u < \infty$ with finite positive moments given by
\begin{equation}\label{eq:fnmoments}
f_n = \int_0^\infty u^n \, \mathrm{d}\phi(u),
\end{equation}
for $n = 0,1,2,\dots$. A necessary and sufficient condition~\cite{Baker1996pade} for a function to be Stieltjes is that all the determinants $\mathcal{D}_{m,n}^{(f)}$ given by
\begin{equation}
\mathcal{D}_{m,n}^{(f)} =
\begin{vmatrix}
f_m & f_{m+1} & \cdots & f_{m+n} \\
f_{m+1} & f_{m+2} & \cdots & f_{m+n+1} \\
\vdots & \vdots & & \vdots \\
f_{m+n} & f_{m+n+1} & \cdots & f_{m+2n}
\end{vmatrix},
\label{eq:determinant}
\end{equation}
are positive for any $m \geq 0$ and $n \geq 0$. These determinants produce constraints between the Taylor series coefficients $f_n$ of the Stieltjes functions. They will also prevent the appearance of \textit{defects}, or technically Froissart doublets, exact cancellations between the zeros of the numerator and the denominator~\cite{Baker1996pade}, effectively reducing the order of the approximants. These will be essential in our analysis since a PA or D-Log constructed from a Stieltjes function has the same Stieltjes properties as the original function has~\cite{Baker1975essentials,Baker1996pade}.
The D-Log in Eq.~(\ref{eq:dlogpa}) is, in general, not a rational approximant, but the function $F(z)$ is meromorphic and, because of that, it is easily approximated by the PA $\bar P_M^N$, and the convergence of the approximation can be proven~\cite{Baker1975essentials}. What is more, for some Stieltjes functions $f(z)$, it can be shown that $F(z)$ will also be Stieltjes. In these cases, convergence theorems for Stieltjes functions apply to the approximation of $F(z)$ by $\bar P_M^N$ as well.
The exception to the convergence rule is the $P^N_1$ sequence as one can prove that approximants belonging to this sequence are not necessarily Stieltjes even when built to approximate a Stieltjes function~\cite{Baker1996pade}.

It is possible to prove~\cite{Masjuan:2009wy,Aubin:2012me} that $\Delta\alpha_{\mathrm{had}}$, which is related to the inclusive hadronic electroproduction cross-section, is a Stieltjes function defined in the region $-\infty < t \leq 0$. Thus, Pad\'e Theory~\cite{Baker1975essentials,Baker1996pade} assures that the poles of approximants of the type $P_M^{M+ k}$ with $k\geq-1$, as well as the branch point of D-Logs of the type $D_N^N$, $D_N^{N+1}$ and $D^N_{N+1}$ to $\Delta\alpha_{\mathrm{had}}$, are always real and positive with positive residues. Furthermore, since $\Delta\alpha_{\mathrm{had}}$ scales as ${\cal O}(t^1)$ for small $t$, the PA sequences $P_N^N$ and both $P_N^{N+1}$ and $P^N_{N+1}$ bound the original function. In our case, the fastest convergence is obtained with the $P_N^{N+1}$, then the convergence theorem for Stieltjes functions reads
\begin{equation}
P_1^1(t) \leq P_2^2(t) \leq \dots \leq \Delta\alpha_{\mathrm{had}} \leq \dots \leq P_2^3(t)\leq P_1^2(t) \, .
\label{eq:convergence}
\end{equation}
\noindent
For the D-Log case, a similar pattern is found, in our case. The sequence $D_N^N$ together with $D^N_{N+1}$ and $D_N^{N+1}$ bound the original function. We found the fastest convergence with the $D^N_{N+1}$ sequence, following the pattern
\begin{equation}
D_1^1(t) \leq D_2^2(t) \leq \dots \leq \Delta\alpha_{\mathrm{had}} \leq \dots \leq D_3^2(t) \leq D_2^1(t).
\label{eq:convergenceDlogs}
\end{equation}

Besides being model-independent, this method takes advantage of the convergence theorems of Pad\'e Theory. To construct the functions that will be used to fit the MUonE toy data in a form that makes contact with the convergence theorems, we will first compute the PAs and D-Logs to the (unknown) Taylor series of $\Delta\alpha_{\mathrm{had}}(t)$, i.e., we will build canonical approximants to
\begin{equation}
\Delta\alpha_{\mathrm{had}}(t) = a_1\,t + a_2\,t^2 + a_3\,t^3 + \dots ,
\label{eq:seriesdeltaalpha}
\end{equation}
where the coefficients $a_n$ are unknown. Then, after a change of variable from $t$ to $x$ using Eq.~(\ref{eq:ttox}), we finally get our fitting functions as a function of $x$. With this technique, we can use our knowledge of Stieltjes functions in full to analyze the fit quality and provide constraints to the fit parameters.

In particular, from Carleman's condition~\cite{Baker1996pade,carleman1923approximation}, the Taylor coefficients of a Stieltjes series cannot change sign and, in our case, they have to be negative. Moreover, the determinant condition imposes the following hierarchy for the coefficients
\begin{equation}
0 > a_i > a_{i+1}, \qquad \quad i \in \mathbb{N}.
\label{eq:coeffhierarchy}
\end{equation}
All these constraints will be used in the fit procedure and it is important to mention that they are model-independent, relying only on $\Delta\alpha_{\mathrm{had}}(t)$ being a Stieltjes function.

Finally, we mention that, in our case, since $\Delta\alpha_{\mathrm{had}}(t)$ has $a_0=0$, the $P^N_M(z)$ will match $N+M$ coefficients instead of $N+M+1$, and the $D_M^N(z)$ $N+M$ instead of $N+M+2$.

\section{A model for the Euclidean correlator}\label{sec:model}

\begin{figure}[t]
\centering
\includegraphics[width=0.5\textwidth]{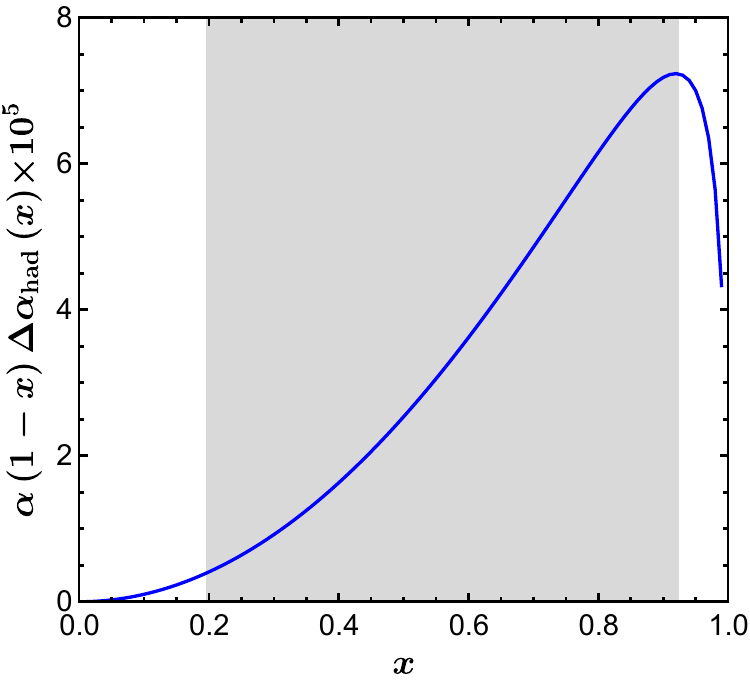}
\caption{The integrand to obtain $a_\mu^{\mathrm{HVP, \,LO}}$, given in Eq.~(\ref{eq:amu}), calculated from the model of Greynat and de Rafael of Eq.~(\ref{eq:model}). The gray area is the expected region that the MUonE experiment will cover.}
\label{fig:model}
\end{figure}

To test our method, we need to generate toy data sets based on a sufficiently realistic model. We will use the phenomenological model for the function $\operatorname{Im} \Pi_{\mathrm{had}}(s)$ inspired by chiral perturbation theory and perturbative QCD introduced by Greynat and de Rafael in Ref.~\cite{Greynat:2022geu} to obtain $\Delta\alpha_{\rm had}$ through the dispersion relation in Eq.~(\ref{eq:dispersiverelation}) and Eq.~(\ref{eq:Deltaalpha}). The model is given by
\begin{equation}
\operatorname{Im} \Pi_{\mathrm{had}} (s) = \dfrac{1}{4\pi} \left( 1 - \dfrac{4 m_\pi^2}{s} \right)^{3/2} \left( \dfrac{|F(s)|^2}{12} + \sum_f Q_f^2 \,\, \Theta (s, s_c, \Delta) \right) \theta (s - 4m_\pi^2),
\label{eq:model}
\end{equation}
with $Q_f$ being the electric charge of the quark of flavor $f$ and $\theta(x)$ the Heaviside theta function. The function $|F(s)|^2$ is the pion vector form factor, which is modeled simply by the $\rho(770)$ contribution as
\begin{equation}
|F(s)|^2 = \dfrac{m_\rho^4}{(m_\rho^2 - s)^2 + m_\rho^2 \, \Gamma(s)^2} ,
\label{eq:functionFrho}
\end{equation}
where the running width is
\begin{equation}
\Gamma(s) = \dfrac{m_\rho \, s}{96 \pi f_\pi^2} \left[ \left( 1 - \dfrac{4 m_\pi^2}{s} \right)^{3/2} \theta(s - 4m_\pi^2) + \dfrac{1}{2} \left( 1 - \dfrac{4 m_K^2}{s} \right)^{3/2} \theta(s - 4m_K^2) \right] ,
\end{equation}
with $f_\pi = 93.3 \, \mathrm{MeV}$ being the pion decay constant and $m_{\pi/K/\rho}$ the corresponding meson masses. The function $\Theta(s,s_c,\Delta)$ is defined as
\begin{equation}
\Theta(s,s_c,\Delta) =
\left[
\dfrac{\arctan{\left(\frac{s-s_c}{\Delta}\right)} - \arctan{\left(\frac{4 m_\pi^2-s_c}{\Delta}\right)}}{\frac{\pi}{2} - \arctan{\left(\frac{4 m_\pi^2-s_c}{\Delta}\right)}} \right].
\end{equation}
The parameters $s_c$ and $\Delta$ will assume the same values employed in Ref.~\cite{Greynat:2022geu}: $s_c = 1\,\mathrm{GeV}^2$ and $\Delta = 0.5 \,\mathrm{GeV}^2$. We show in Fig.~\ref{fig:model} the line shape of the integrand of Eq.~(\ref{eq:amu}) obtained from the use of this model where the gray band represents the experimentally accessible region with the designed MUonE experiment~\cite{CarloniCalame:2015obs,Abbiendi:2016xup,Abbiendi:2677471,Ballerini:2019zkk,Abbiendi:2019qtw,Abbiendi:2021xsh,Abbiendi:2022oks}.

We consider this simple model to be sufficiently realistic because it leads to a representation of $\Delta \alpha_{\mathrm{had}}$ which is a Stieltjes function, as expected in QCD.
Hence, the theorems given in Sec.~\ref{sec:theorypas} and the coefficient constraints of Eq.~(\ref{eq:coeffhierarchy}) are all valid. We also note that numerical checks of several hundred determinants of Eq.~(\ref{eq:determinant}) built from the series of $\frac{d}{dt}\ln\Delta\alpha_{\rm had}(t)$ indicate that this function, which is required to construct the D-Logs, is also Stieltjes.

Computing the integral of Eq.~(\ref{eq:amu}) using the model in Eq.~(\ref{eq:model}) we determine the value of $a_\mu^{\mathrm{HVP,\,LO}}$ from this model as
\begin{equation}
a_{\mu,\,\mathrm{model}}^{\rm HVP,\,LO} = 6992.4 \times 10^{-11}.
\label{eq:amumodel}
\end{equation}
This result will serve as a guide for us to compare the values of $a_\mu^{\mathrm{HVP, \, LO}}$ obtained from our PAs and D-Log approximants.

\section{Approximants to the Taylor series}\label{sec:pastaylor}

In this section, we will canonically build PAs and D-Logs by matching each approximant's Taylor series to that of $\Delta\alpha_{\mathrm{had}}(t)$, Eq.~(\ref{eq:seriesdeltaalpha}), which will be computed from the model of Eq.~(\ref{eq:model}). In this case, the convergence theorems to Stieltjes functions apply. Thus the convergence to the $a_\mu^{\mathrm{HVP,\,LO}}$ value of the model, Eq.~(\ref{eq:amumodel}), is guaranteed and it should respect the pattern determined in Eqs.~(\ref{eq:convergence}) and (\ref{eq:convergenceDlogs}).

The Taylor series of $\Delta\alpha_{\mathrm{had}}$ as a function of $t$ according to the model of Sec.~\ref{sec:model} is
\begin{equation}
\alpha \, \Delta\alpha_{\mathrm{had}}(t) \times 10^5 = -918 \, t - 1752 \, t^2 - 6066 \, t^3 - 31589 \, t^4 -
214058 \, t^5 + \mathcal{O}(t^6).
\label{eq:seriesdeltaalphamodel}
\end{equation}

\subsection{Convergence of PAs results}

\begin{figure}[t]
\centering
\subfloat[]{\label{fig:pastaylor}{\includegraphics[width=0.44\textwidth]{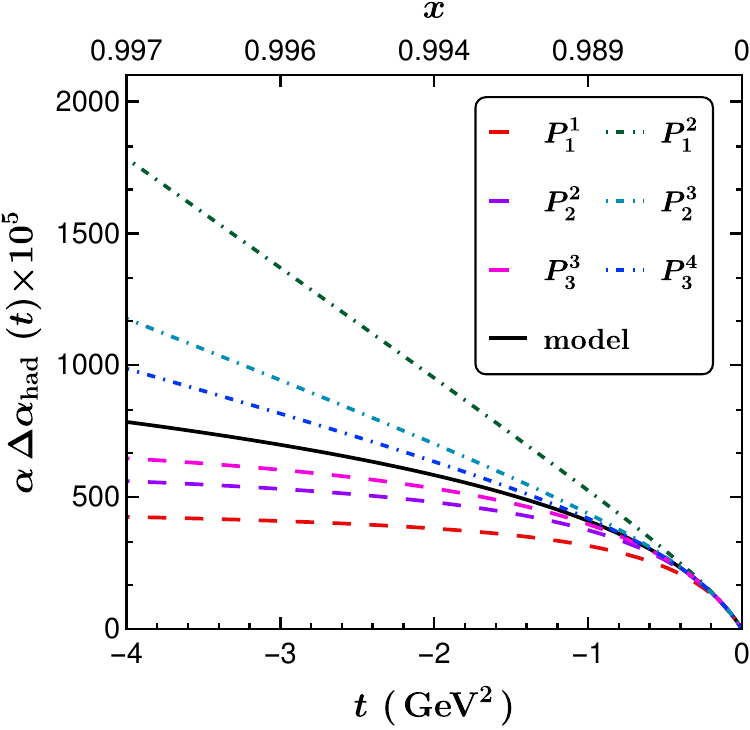}}}\hfill
\subfloat[]{\label{fig:intpastaylor}{\includegraphics[width=0.44\textwidth]{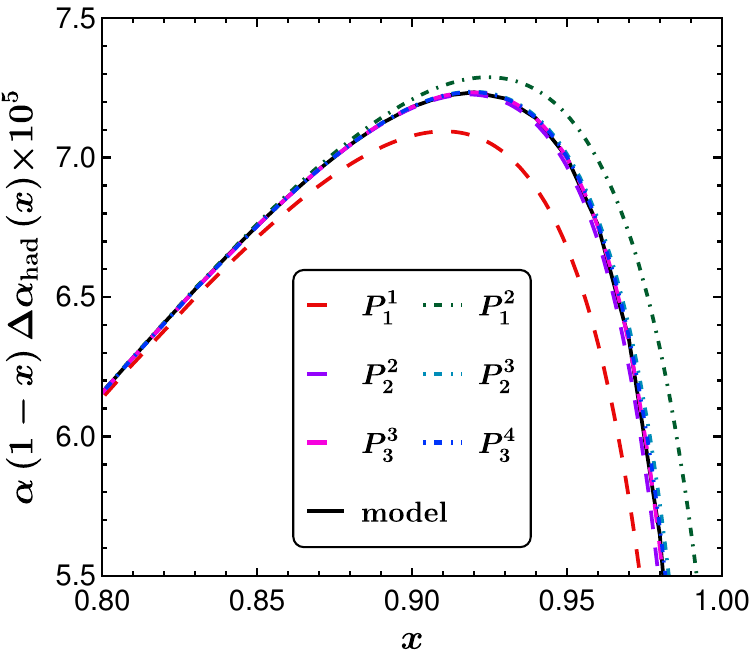}}}
\caption{PAs $P_N^N(t)$ (dashed line) and $P_N^{N+1}(t)$ (dot-dashed line) built from the Taylor series of $\Delta\alpha_{\mathrm{had}}(t)$, Eq.~(\ref{eq:seriesdeltaalphamodel}), together with the model results (black line). (a) $\alpha\Delta\alpha_{\mathrm{had}}(t)$ and (b) $\alpha(1-x)\Delta\alpha_{\mathrm{had}}(x)$, used to compute $a_\mu^{\mathrm{HVP,\,LO}}$ with Eq.~(\ref{eq:amu}).}
\end{figure}

Results from the sequences $P_N^N(t)$ and $P_N^{N+1}(t)$ are shown in Fig.~\ref{fig:pastaylor}, in the range $0\leq x \leq 0.997$, which corresponds to $-4~{\rm GeV}^2\leq t \leq 0~{\rm GeV}^2$. The value from the model for $\Delta\alpha_{\mathrm{had}}$ is represented by the black line in Fig.~\ref{fig:pastaylor}. One can notice that both sequences approach consistently the true value of the model when the order of the PAs is increased and they also bound the model value as expected by the convergence theorem of Eq.~(\ref{eq:convergence}): the PAs of the sequence $P_N^N$ approach the function from below while the ones belonging to the sequence $P_N^{N+1}$ do so from above. Additionally, it is possible to observe that for large values of $|t|$, the discrepancy between the approximants and the model is much more noticeable. Since the PAs are constructed from the Taylor series coefficients of $\Delta\alpha_{\mathrm{had}}(t)$, i.e. in the region of $t$ close to zero, it is to be expected that far from this region the PAs start to deviate from the true model value. Notice, however, that the PA convergence is dramatically faster compared with that of the Taylor expansion, which breaks off outside a radius of convergence around $|t|=4m_{\pi}^2 \sim 0.1$GeV$^2$. The deviation of our approximants at large $t$ is mitigated in the case of the integrand that appears in the calculation of $a_\mu^{\mathrm{HVP,\,LO}}$, which can be seen in Fig.~\ref{fig:intpastaylor} as a function of $x$. Since the integrand goes to zero as $x$ goes to 1, differences in the deep Euclidean region are suppressed, although still noticeable for lower-order PAs. One should also note that the change of variables from $t$ to $x$ maps the entire infinite interval $t\in (-\infty,-4~\rm GeV^2]$ into the small interval $x \in [0.997,1]$.

\begin{figure}[!t]
\centering \includegraphics[width=0.45\textwidth]{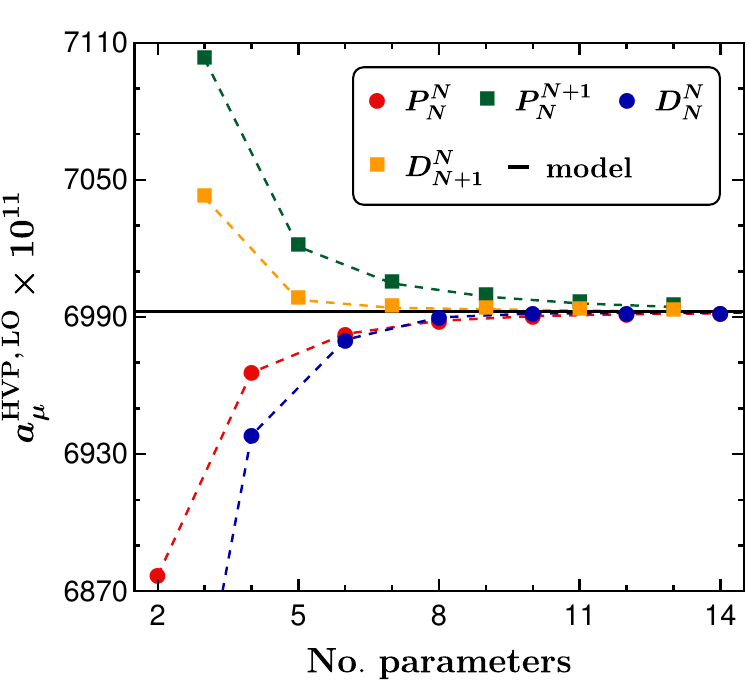}
\caption{Comparison between PA and D-Log estimates of $a_\mu^{\mathrm{HVP,\,LO}}$ and the value predicted by the model given in Eq.~(\ref{eq:amumodel}) (black solid line). `No. parameters' refers to the value of $N+M$ used for each approximant.}
\label{fig:amutaylor}
\end{figure}

We know from Pad\'e Theory~\cite{Baker1996pade, MasjuanQueralt:2010hav,Costin:2021bay} that the approximants can mimic branch cuts by accumulating interleaved poles and zeros along the cut. For higher-order PAs the mimicking of the cut originated by the term $\left(1 - 4 m_\pi^2/t \right)^{3/2}$ in the model of Eq.~(\ref{eq:model}), and whose branch point is located at $t = 0.078 \, \mathrm{GeV}^2$, was observed. Let us take $P_5^6(t)$ as an example:
this PA has poles at $0.089 \, \mathrm{GeV}^2$, $0.116 \, \mathrm{GeV}^2$, $0.187 \, \mathrm{GeV}^2$, $0.426 \, \mathrm{GeV}^2$ and $1.192 \, \mathrm{GeV}^2$, which are interleaved with zeros at $0.089 \, \mathrm{GeV}^2$, $0.118 \, \mathrm{GeV}^2$, $0.198 \, \mathrm{GeV}^2$, $0.620 \, \mathrm{GeV}^2$ and $7.951 \, \mathrm{GeV}^2$.

After building the PAs we can turn to the estimate of the value for $a_\mu^{\mathrm{HVP,\,LO}}$. The results can be seen in Fig.~\ref{fig:amutaylor}, with the black solid line representing the model's value given in Eq.~(\ref{eq:amumodel}). Again, the pattern stated by the theorem in Eq.~(\ref{eq:convergence}) is obeyed, with the PAs sequence $P_N^N(t)$ (illustrated by the dashed red line) reaching the model result from below and sequence $P_N^{N+1}(t)$ (dashed green line) approaching from above. The convergence to the true value, which is guaranteed by theorems, can also be observed.

\subsection{Convergence of D-Logs and comparison with PAs results}

\begin{figure}[t]
\centering
\subfloat[]{\label{fig:dlogstaylor}{\includegraphics[width=0.44\textwidth]{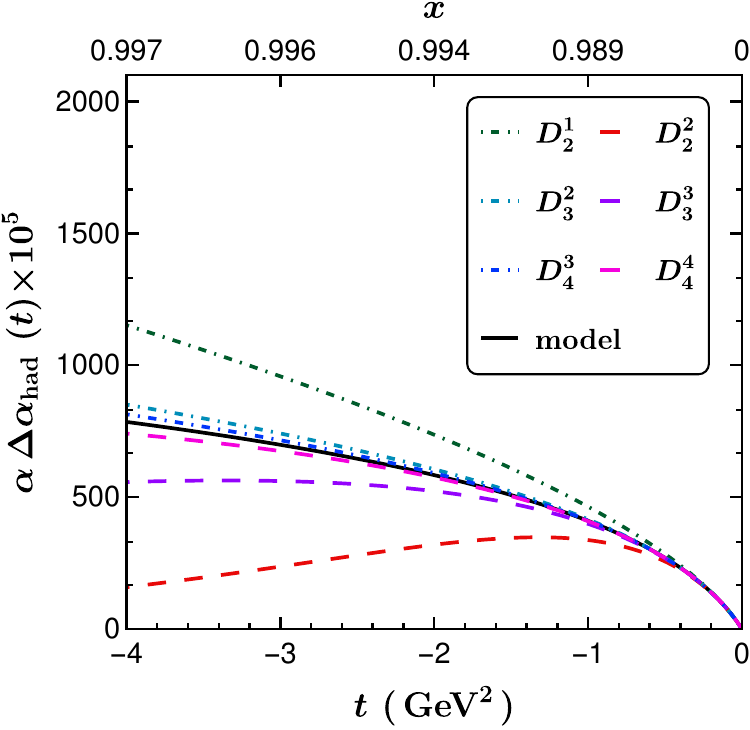}}}\hfill
\subfloat[]{\label{fig:intdlogstaylor}{\includegraphics[width=0.44\textwidth]{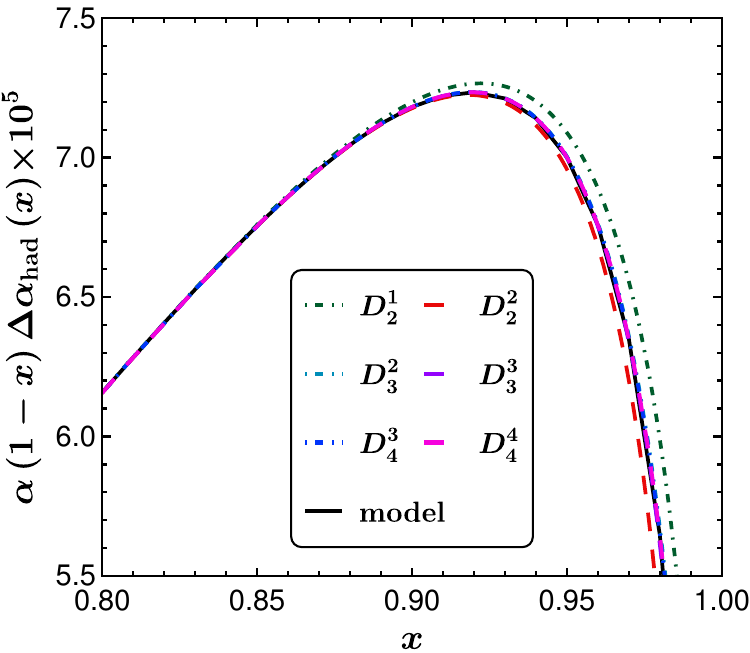}}}
\caption{D-Logs $D_N^N(t)$ (dashed line) and $D_{N+1}^N(t)$ (dot-dashed line) built from the Taylor series of $\Delta\alpha_{\mathrm{had}}(t)$, Eq.~(\ref{eq:seriesdeltaalphamodel}), together with the model results (black line). (a) $\alpha\Delta\alpha_{\mathrm{had}}(t)$ and (b) $\alpha (1-x) \Delta\alpha_{\mathrm{had}}(x)$. Plots in the same scale as Figs.~\ref{fig:pastaylor} and~\ref{fig:intpastaylor}.}\label{Fig:DLogconvergence}
\end{figure}

We can also build D-Logs to the Taylor series of $\Delta\alpha_{\mathrm{had}}(t)$. As stated by the convergence theorems for Stieltjes functions, specific sequences of D-Logs to $\Delta\alpha_{\mathrm{had}}(t)$ bound the function from above and from below, as can be seen in Fig.~\ref{Fig:DLogconvergence}, in complete analogy to the convergence pattern of PAs shown in Figs.~\ref{fig:pastaylor} and~\ref{fig:intpastaylor}.

Compared to PAs, D-Logs contain built-in branch cuts. For example, $D^5_5(t)$ has $4$ branch cuts and $1$ exponential factor. The onset of the first branch cut for each D-Log approaches the two-pion production threshold of the model, and this is done hierarchically: the D-Log $D^3_3(t)$ predicts the production threshold to be at $0.095 \, \mathrm{GeV}^2$ and the D-Log $D^9_9(t)$ at $0.0798 \, \mathrm{GeV}^2$. With this value, we can predict the charged pion mass to be 0.141~GeV, to compare with 0.140~GeV used in the model of Eq.~(\ref{eq:model}).
We also noticed all D-Logs have a branch cut that tends to be a square root with the singularity around $0.5 \, \mathrm{GeV}^2$. This branch cut can be interpreted as a signal of the $\rho$ meson mass and an attempt to replicate the Breit-Wigner distribution used by the model in \cref{eq:functionFrho}.

Finally, in Fig.~\ref{fig:amutaylor} we compare the performance of both PAs and D-Logs provided the same amount of coefficients of the Taylor expansion are used. Subdiagonal $D_{N+1}^N$ performs better than $P^{N+1}_N$ when approaching from above for all orders, while diagonal PAs do better than diagonal D-Logs when approaching from below. Notice however that the latter comparison holds up only to the number of parameters up to $6$, above which the diagonal D-Logs surpass PAs as they converge faster. Not only Cauchy convergence is observed in both sequences but absolute convergence as well. Both methods are useful when bounding the function. For a large number of parameters, D-Logs would converge faster but for lower orders, none of the methods is systematically superior. We shall keep both methods for our fitting function study.

\section{Approximants to data with no error}\label{sec:pasnoerror}
Let us consider now an idealized case, where the data points represent exactly what is expected from the model of Eq.~(\ref{eq:model}), with zero error. It is convenient to work with data for $\alpha \, \Delta\alpha_{\mathrm{had}}(x) \times 10^5$ such that the Taylor coefficients $a_n$ are of a natural size, which makes the fits simpler.
The values of $x$ were calculated as follows: the interval $0.2\leq x\leq 0.93$ was divided into 30 equally spaced bins and we take the center of each bin, $x_i$, without any error, as the representative values of $x$ where we will generate the data points to be fitted with our approximants.

To obtain our fitting functions, both PAs and D-Logs will be written in terms of the Taylor series coefficients of $\Delta\alpha_{\mathrm{had}}(t)$ and later converted to the $x$ variable by Eq.~(\ref{eq:ttox}). These functions will then be used to perform the fits to the toy data sets where the fit parameters can be written in terms of the Taylor coefficients of $\Delta\alpha_{\mathrm{had}}(t)$. PAs used as fitting functions were already applied in similar contexts and the convergence theorems are apparently satisfied in these cases~\cite{Masjuan:2008fv,Masjuan:2012wy,Masjuan:2013jha, Escribano:2013kba,Escribano:2015yup,Masjuan:2017tvw}. It is important to emphasize, however, that the theorems presented in Sec.~\ref{sec:theorypas} are demonstrated in the case of canonical PAs, i.e. those built to the Taylor series coefficients. Strictly speaking, in the case of a fit to data points in a given interval, we conscientiously slightly depart from the conditions of the theorem as a fit to data can be interpreted as imposing one matching condition for each datum instead of $N+M$ conditions at the same point. Therefore, we expect convergence theorems to be, as we will see, satisfied in this case as well but the convergence velocity may be smaller.

To show in detail a concrete case of our procedure, we start by examining the PA $P_1^1(t)$. To construct the fitting function, we first compute $P_1^1(t)$ as a function of the unknown Taylor series coefficients $a_n$ of $\Delta\alpha_{\mathrm{had}}(t)$ given by Eq.~(\ref{eq:seriesdeltaalpha}). We Taylor-expand the PA and $\Delta\alpha_{\mathrm{had}}(t)$ and match the coefficients of both order by order. We then perform the change of variables of Eq.~(\ref{eq:ttox}) which leads to the fitting function in terms of $x$\footnote{We will use the same notation $P_M^N(x)$ to refer to the fitting functions derived from the PA $P_M^N(t)$ even though they are not a PA of order $N+M$ in the variable $x$.}
\begin{equation}
P_1^1(t(x)) \equiv P_1^1(x) = - \dfrac{a_1^2 \, m_\mu^2 \, x^2}{a_1 - a_1 \, x + a_2 \, m_\mu^2 \, x^2} = - \dfrac{b_1 \, m_\mu^2 \, x^2}{1 - x + b_2 \, m_\mu^2 \, x^2} \,,
\end{equation}
where the fit parameters are $b_1 = a_1$ and {$b_2 = a_2/a_1$}. Due to the hierarchy of the Taylor coefficients of $\Delta\alpha_{\mathrm{had}}(t)$, given in Eq.~(\ref{eq:coeffhierarchy}), we can conclude that $b_1$ and $b_2$ have to obey the following relations: $b_1<0$ and $b_2>1$, which will be imposed in the fit.

To obtain the parameters of the PAs and D-Logs from these fits to this zero-error data, we will perform the minimization of the fit quality $\mathcal{Q}^2$ given simply by
\begin{equation}
\mathcal{Q}^2 = \sum_{i=1}^{30} \left[ \alpha \, \Delta\alpha_{\mathrm{had}}(x_i) \times 10^5 - \mathcal{P}_M^N(x_i) \right]^2,
\end{equation}
where $\mathcal{P}_M^N(x)$ refers, generically, to a PA $P_M^N(x)$ or to a D-Log $D^N_M(x)$.
In reasonable fits, we expect very small values for $\mathcal{Q}^2$. For $P^1_1(x)$, after the minimization we get $\mathcal{Q}^2 = 1.18 \times 10^{-3}$ and the values of the fit parameters $b_1$ and $b_2$ lead to the following Taylor series coefficients
\begin{equation}
a_1 = -912 \, \mathrm{GeV}^{-2} \qquad \mathrm{and} \qquad a_2 = -1489 \, \mathrm{GeV}^{-4},
\end{equation}
which differ from the true values by only 0.7\% and 15\%, respectively. Furthermore, the pole of this PA in $t$ is at $t_{\mathrm{pole}} = 0.61 \, \mathrm{GeV}^2$. As stated by the convergence theorems, the pole is located on the positive real axis of $t$.

The PA can then be used to estimate the value of $a_\mu^{\mathrm{HVP,\,LO}}$. By employing the resulting $P^1_1(x)$ in Eq.~(\ref{eq:amu}) we get
\begin{equation}
a_{\mu,\,P^1_1}^{\mathrm{HVP,\,LO}} = 6933 \times 10^{-11},
\end{equation}
with an error of 0.9\% compared to Eq.~(\ref{eq:amumodel}), the expected value within our model.

\begin{figure}[t]
\centering
\subfloat[]{\label{fig:amupasnoerror}{\includegraphics[width=0.48\textwidth]{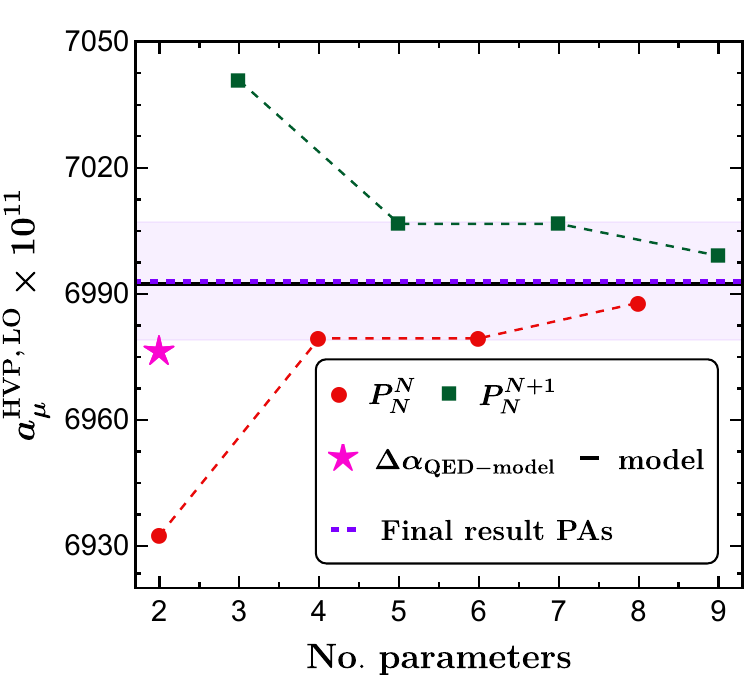}}}\hfill
\subfloat[]{\label{fig:amunoerrorDlogs}{\includegraphics[width=0.48\textwidth]{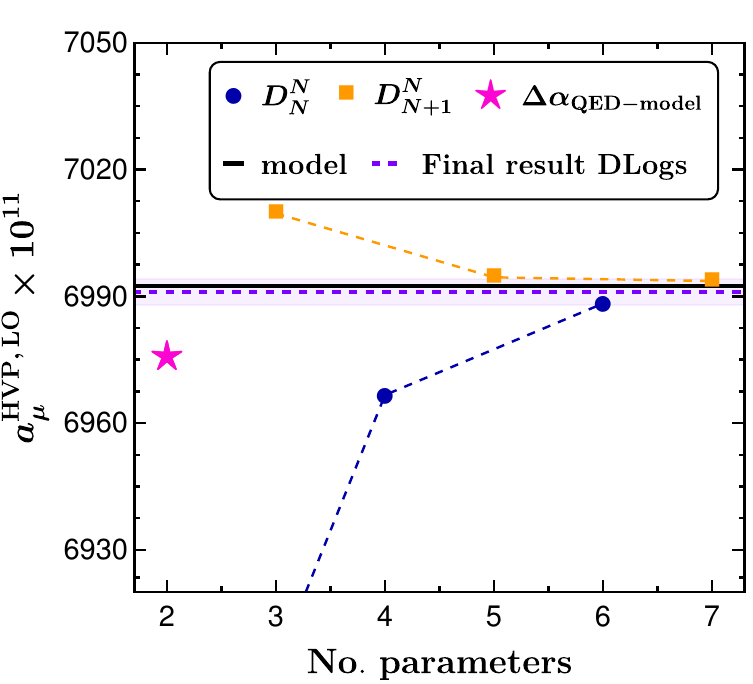}}}
\caption{$a_\mu^{\mathrm{HVP,\,LO}}$ from (a) PAs and (b) D-Logs fitted to a data set with zero error.
The pink star is the value obtained from the model of Eq.~(\ref{eq:deltaalphaqed}) and the purple dashed line is the final result coming from the (a) PAs and (b) D-Logs with the light-purple band representing the systematic error of the result. The black line indicates the true value of the model given in Eq.~(\ref{eq:amumodel}).}
\end{figure}

The same analysis can be made for the other PAs of the sequences $P_N^N(t)$ and $P_N^{N+1}(t)$.\footnote{The fitting functions obtained from these approximants are given explicitly in Appendix~\ref{app:pasfunctions}.} We will employ only these two sequences since they are expected to bound the real function in the theorem presented in Sec.~\ref{sec:theorypas}. We will build several approximants from each sequence, in order to study the convergence pattern, but the final estimate of $a_\mu^{\mathrm{HVP,\,LO}}$ will not include all results, as we detail below.
Fig.~\ref{fig:amupasnoerror} shows the values predicted by the PAs of the sequences $P_N^N$ and $P^N_{N+1}$ for $a_\mu^{\mathrm{HVP,\,LO}}$. It is possible to notice that the pattern of convergence of Eq.~(\ref{eq:convergence}), where the $P_N^N$ bound the true value from below and the ones of $P^N_{N+1}$ from above, is satisfied in this case. One can note in Fig.~\ref{fig:amupasnoerror} that in some cases when the order of the PA is increased there is no real improvement in the estimate of $a_\mu^{\mathrm{HVP,\,LO}}$. This can be explained by the appearance of a \textit{defect}, which consists of a pole partially canceled by a nearby zero, effectively reducing the order of the PA. It is known from PA theory~\cite{Baker1975essentials,Baker1996pade} that approximants applied canonically to Stieltjes functions, i.e. to their Taylor series, cannot have Froissart doublets. Nonetheless, the theorem does not prevent the PA to have numerical defects, which are almost exact cancellations between poles and zeros, arising from the fitted parameters. Adding more coefficients into a fitting function also introduces larger correlations among them. As such, fitted coefficients will have errors and within their allowed fitted regions, cancellations may emerge. We note, however, that the important fact that the two sequences of PAs bound the true value remains true.

A final estimate for $a_\mu^{\mathrm{HVP, \, LO}}$ from the PAs can then be obtained. For that, we will limit the analysis to the first two approximants in each sequence, i.e. $P_1^1$, $P_2^2$, $P_1^2$ and $P_2^3$. This choice is motivated by the fact that we do not expect to be able to fit much more than 5 parameters in more realistic scenarios, with a limited number of data points and with the errors expected by the MUonE experiment. Computing the central value of the prediction as the mean between the four approximants we get $a_{\mu,\,\mathrm{PAs}}^{\mathrm{HVP, \, LO}} = 6990 \times 10^{-11}$, which differs from the expected value given in Eq.~(\ref{eq:amumodel}) by $0.04\%$. This can be improved if we take advantage of the expected convergence of the PA sequences dictated by Eq.~(\ref{eq:convergence}). It is expected that the higher-order PAs will be closer to the true value. We can then take the average of the highest-order approximants of each sequence only, i.e. $P_2^2$ and $P_2^3$, since from the convergence behavior of Eq.~(\ref{eq:convergence}), we expect the true value to lie between these two PAs estimates. Thus, computing this mean we get $a_{\mu,\,\mathrm{PAs}}^{\mathrm{HVP, \, LO}} = 6993 \times 10^{-11}$, that presents an error of less than $0.01\%$ with respect to the model value. This result can be seen in Fig.~\ref{fig:amupasnoerror} as the purple dashed line.

One can also estimate a theoretical uncertainty due to the truncation of the PAs sequences. Taking the convergence pattern of Eq.~(\ref{eq:convergence}) into consideration, the error will be defined as half of the distance between the two highest-order PAs ($P_2^2$ and $P_2^3$ in this case), which gives us a relative error of 0.19\% represented in Fig.~\ref{fig:amupasnoerror} by the light-purple band. 
This systematic error has to be added to our final numbers in the spirit of a conservative estimate.

A similar analysis can be performed with D-Logs. As argued in the previous section, they introduce branch cuts thus speeding the convergence, a key factor when dealing with a small number of parameters. In this case, we will use the approximants of the sequences $D_N^N$ and $D^N_{N+1}$, since these are the sequences expected to bound the original function, as indicated in Eq.~(\ref{eq:convergenceDlogs}). To construct the fitting functions, we start with an arbitrary Taylor series in the $t$ variable up to order $N+M$, compute its logarithm and then the derivative with respect to $t$, as indicated in Eq.~(\ref{eq:Fdlog}).
After that, we construct a PA, $\bar{P}_M^N$, to the thus obtained series and, employing Eq.~(\ref{eq:dlogpa}), we obtain the D-Log expression in terms of $t$. We then perform the change of variables of Eq.~(\ref{eq:ttox}) and write the generic fitting function in terms of $x$.
The simplest D-Log, for example, is $D_2^1(t)$ that before and after the change of variables reads
\begin{equation}
D_2^1(t)= \frac{-f_0\,t}{(r_1-t)^{\gamma_1}}\, \rightarrow D_2^1(x)=\frac{f_0\,m_{\mu}^2\,x^2 (1-x)^{-1+\gamma_1}}{(r_1-r_1x+m_{\mu}^2x^2)^{\gamma_1}},
\end{equation}
with fit parameters $f_0$, $r_1$ and $\gamma_1$. We recall that $D_M^N$ requires $N+M$ fit parameters.
The next approximant is $D_2^2(t)$, which is given by
\begin{equation}\label{eq:D22}
D_2^2(t)= \frac{-f_0\,t\,e^{\beta t}}{(r_1-t)^{\gamma_1}}\, \rightarrow
D_2^2(x)=\frac{f_0\,m_{\mu}^2\,x^2 (1-x)^{-1+\gamma_1}}{(r_1-r_1x+m_{\mu}^2x^2)^{\gamma_1}}e^{\beta\frac{m_{\mu}^2x^2}{(x-1)}}\, ,
\end{equation}
with fit parameters $f_0$, $r_1$, $\gamma_1$ and $\beta$. All D-Logs used are provided in Tab.~\ref{tab:dlogtypefunction} in Appendix~\ref{app:pasfunctions}.

A comparison of the results for $a_\mu^{\rm HVP,\,LO}$ obtained from the D-Log approximants is found in Fig.~\ref{fig:amunoerrorDlogs} where we show both sequences, $D_N^N$ and $D^N_{N+1}$ as a function of the number, $N+M$, of fitted coefficients. Convergence for D-Logs is faster than PAs, especially for the subdiagonal sequence. In line with the criteria used for PAs, we employ $D^2_3$ and $D^3_3$ to obtain a final prediction for $a_{\mu}^{\mathrm{HVP,\,LO}}$ from the D-Logs. Computing the mean of these values, we obtain $a_{\mu,\,\mathrm{DLogs}}^{\mathrm{HVP, \, LO}} = 6991.4 \times 10^{-11}$, which exhibits an error of less than $0.02\%$ compared to the model value and a theoretical uncertainty band of less than $0.05\%$ (light-purple band in \cref{fig:amunoerrorDlogs}).

An important outcome of analyzing data with no errors is to provide an estimate of the theoretical uncertainty for other models proposed in the literature to fit and extrapolate the MUonE data. Let us take a closer look at the model proposed by Abbiendi in the letter of intent of the MUonE experiment, Ref.~\cite{Abbiendi:2677471}, motivated by the one-loop QED calculation of the vacuum polarization. This fitting function is
\begin{equation}
\Delta\alpha_{\mathrm{QED-model}}(t) = K M \left[ -\frac{5}{9} - \dfrac{4M}{3t} + \dfrac{2\left(\frac{4M^2}{3t^2} + \frac{M}{3t} - \frac{1}{6} \right)}{\sqrt{1-\frac{4M}{t}}} \log \left| \dfrac{1-\sqrt{1-\frac{4M}{t}}}{1+\sqrt{1-\frac{4M}{t}}} \right| \right],
\label{eq:deltaalphaqed}
\end{equation}
where the parameters $K$ and $M$ are determined by the fit.

After the $\mathcal{Q}^2$ minimization we obtain $\mathcal{Q}^2 = 1.84 \times 10^{-4}$ and the parameters determined by the fit are: $K = 6865.36 \,\, \mathrm{GeV}^{-2}$ and $M = 0.06 \,\, \mathrm{GeV}^2$. The $a_\mu^{\mathrm{HVP,\,LO}}$ value can be predicted by this model employing the expression above in Eq.~(\ref{eq:amu}), which in this scenario gives us $a_{\mu,\,\mathrm{QED-model}}^{\mathrm{HVP, \, LO}} = 6976 \times 10^{-11}$. Even though this function has only two parameters, as is the case of $P_1^1(x)$, this model implements the logarithmic dependency expected at large $|t|$, which facilitates a better approximation to the exact result. Thus, it is expected that $\Delta\alpha_{\mathrm{QED-model}}(t)$ has a more accurate result than $P_1^1(x)$, which can be verified in Fig.~\ref{fig:amupasnoerror} where the pink star represents the prediction of $\Delta\alpha_{\mathrm{QED-model}}$. Regarding the D-Logs, $D_1^1(t)$, which contains only two parameters, is unable to reproduce any type of singularity, whether poles or cuts. Therefore one would not expect that its estimate of $a_\mu^{\mathrm{HVP,\,LO}}$ be more precise than the one provided by the QED-inspired model, which is corroborated by Fig.~\ref{fig:amunoerrorDlogs}. The next D-Log, $D^1_2(t)$, with three parameters, contains a branch cut and achieves a relative error of $0.24\%$ when compared to the true value of our model, Eq.~(\ref{eq:amumodel}), an error very similar to one obtained from $\Delta\alpha_{\mathrm{QED-model}}$, but using our model-independent method. When comparing it, however, to our final estimate from the PAs or D-Logs given in Figs.~\ref{fig:amupasnoerror} and~\ref{fig:amunoerrorDlogs}, respectively, the result from the use of $\Delta\alpha_{\mathrm{QED-model}}$ differs significantly more from the true model value of Eq.~(\ref{eq:amumodel}). A priori knowledge of the function has been used when defining the QED-model in Eq.~(\ref{eq:deltaalphaqed}). Even though the logarithmic dependence is captured, the value of the fitted parameter $M$ departs from the true one, $m_{\pi}^2 \sim 0.02$GeV$^2$. The accuracy of the result is thus model-dependent and the systematic error is difficult to quantify. Whether the QED-model in Eq.~(\ref{eq:deltaalphaqed}) would perform similarly with real data cannot be answered with guarantees.

As mentioned above, we can estimate the theoretical uncertainty in using $\Delta\alpha_{\mathrm{QED-model}}$ as the relative difference between the fitted value and the true model value given in Eq.~(\ref{eq:amumodel}). This gives us $0.24\%$, which is slightly larger than, but of the same order of, the one obtained with our PA method. The systematic error from the use of $\Delta\alpha_{\mathrm{QED-model}}$ is, however, significantly larger than that of the D-Log approach. As already stated, these systematic uncertainties have to be considered in the final estimates of $a_\mu^{\mathrm{HVP,\,LO}}$ regardless of the fitting function used.

\section{Approximants to data with realistic errors}\label{sec:results}

We can now move to more realistic data sets, with fluctuations and uncertainties of the same order as those expected to be obtained in the MUonE experiment. We built 1000 toy data sets for the function $\alpha \, \Delta\alpha_{\mathrm{had}}(x) \times 10^5$ employing the model of Eq.~(\ref{eq:model}) with 30 data points each, corresponding to the expected MUonE bin sizes. The 30 values of $x$ are computed in the same way as in the previous section. The data sets were generated assuming a Gaussian distribution around the value of $\alpha \, \Delta\alpha_{\mathrm{had}}(x) \times 10^5$ given by the model of Eq.~(\ref{eq:model}) with an error ranging from 0.7\%, for larger values of $x$, up to 6.7\% for $x \approx 0.2$.\footnote{The authors thank Giovanni Abbiendi, Carlo Carloni Calame, and Graziano Venanzoni for providing us with the values of the expected uncertainties of the MUonE experiment.} Additionally, the $a_\mu^{\mathrm{HVP,\,LO}}$ of each data set was calculated: the data was used in the region $x \in [0.2,0.93]$, and the model of Sec.~\ref{sec:model}, used in the data generation, was employed outside this interval. Calculating the median and the $68\%$ confidence level (CL) of this distribution we get
\beq
a_\mu^{\mathrm{HVP,\,LO}} = (6991^{+22}_{-20}) \times 10^{-11}\qquad \mbox{(model value)}.
\eeq
Since in this case the extrapolation outside the data region is exact, this value gives an idea of the best possible result that we can expect from our PA or D-Log predictions given the available information in the data sets.

For each data set, the parameters $b_n$ of the PAs and $f_0, \beta, r_n,\gamma_n$ of the D-Logs are determined by a $\chi^2$ minimization. We apply penalizations if the result for the Taylor series coefficients $a_n$ of $\Delta\alpha_{\mathrm{had}}$ do not follow the expected hierarchy given in Eq.~(\ref{eq:coeffhierarchy}) or the determinant condition discussed in Sec.~\ref{sec:theorypas}. This is done by employing a modified $\chi^2$ function given by
\begin{align}
\chi^2(\mathcal{P}_M^N) = \sum_{i,j=1}^{30} &\left[ \alpha \, \Delta\alpha_{\mathrm{had}}(x_i) \times 10^5 - \mathcal{P}_M^N(x_i) \right] (C^{-1})_{ij} \left[ \alpha \, \Delta\alpha_{\mathrm{had}}(x_j) \times 10^5 - \mathcal{P}_M^N(x_j) \right] \nonumber \\
&+ n_{\mathrm{dof}}\left[ \sum_{i=2}^{N+M} \theta{(a_{i} - a_{i-1})} + \sum_{i=3}^{N+M} \theta{(\mathcal{D}_{i-1,1})} + \sum_{i=5}^{N+M} \theta{(-\mathcal{D}_{i-2,2})}\right],
\label{eq:chi2min}
\end{align}
where $C$ is the data covariance matrix, $n_{\mathrm{dof}}$ is the number of degrees of freedom, $\theta(x)$ is the Heaviside theta function and $\mathcal{D}_{m,n}$ is the determinant given in Eq.~(\ref{eq:determinant}). The $\chi^2$ penalties are scaled by $n_{\rm dof}$ to be of a natural size.
Hence, if the hierarchy of $a_i$ coefficients or the determinant conditions are not satisfied, the $\chi^2$ {has a steep increase which forces the minimization algorithm to search for minima respecting the conditions expected for Stieltjes functions. The arguments inside $\theta(x)$ are always written in terms of the fit parameters of each approximant. Alternatively, one could neglect the second line from Eq.~(\ref{eq:chi2min}) and turn the aforementioned penalization into limits for the fit parameters. Both strategies lead to equivalent results.

For each fit, the approximant written as a function of the variable $t$ is examined for the appearance of defects. As already explained in Sec.~\ref{sec:pasnoerror}, there is no guarantee that numerical defects will not occur when approximants are used as fitting functions to real data. In PAs, defects are manifest in a nearly exact cancellation between a pole and a zero of the approximant. Their presence effectively decreases the order of the PA, which spoils the systematic study of the convergence of a given sequence. For the D-Logs, a similar effective reduction in the order of the approximants can happen in several circumstances. If the exponent of a cut, $\gamma_n$, is compatible with zero, this reduces the number of fitted coefficients by two, for example. Equivalently, if two branch points are equal, i.e. $r_i=r_j\, ({i\neq j})$, the two cuts are merged, again lowering the number of coefficients by two. In the case of the diagonal D-Logs, the exponential coefficient, $\beta$, may be compatible with zero within errors, reducing in practice the approximant by one order (see Appendix~\ref{app:pasfunctions} for the explicit D-Log expressions). Finally, if $\sum_n\gamma_n=1$, the pole at $x=1$, resulting from the change of variable described in Eq.~(\ref{eq:ttox}), is lost.

In summary, those approximants with numerical defects are discarded because they are redundant with lower order ones.\footnote{When we inspect the approximants for the appearance of these defects, we employ a numerical tolerance of $10^{-4}$. This number is somewhat arbitrary, and we have checked that varying it by one order of magnitude does not alter the results significantly.} For instance, for the PA $P_2^2$ approximately 30\% of the fits have to be disregarded due to the presence of defects, and for $P_2^3$ this number is 56\%. In the case of D-Logs, $0.7\%$ of fits for $D^2_3$ and $4\%$ for $D^3_3$ are not considered.

A related issue appears in fits where one of the parameters turns out to be almost zero without numerical cancellation. In our fits, this happens when using the PA $P_2^2$ as a fitting function for several of the data sets.\footnote{This only happens for $P^2_2$ and this is not observed when fitting the data without fluctuations, which indicates that this is an artifact related to the statistical errors.} This may be problematic since, with one of the parameters in the denominator equal to zero, this approximant is effectively reduced to $P^2_1$. To circumvent this issue, in our results for $P_2^2$, we impose that none of the parameters are zero, even if this leads to a larger $\chi^2$, which guarantees that we do not mix results from different approximants in the final statistical distributions.

After fitting, we use the resulting approximant to calculate $a_\mu^{\rm HVP,\,LO}$ for each of the accepted fits using Eq.~(\ref{eq:amu}) (or with Eq.~(\ref{eq:amuxmax}), for the partial contributions).\footnote{It is also possible to use the approximants only to extrapolate the data, i.e. use the data points to obtain $a_\mu^{\mathrm{HVP,\,LO}}$ in the region $x \in [0.2, 0.93]$ and then apply the approximants only outside this interval. The results from both procedures are in very good agreement, since the fit quality is almost always excellent, which is confirmed by Fig.~\ref{fig:pasdata} and the $\chi^2$ values of Tab.~\ref{tab:PAsamu}.}
The results from each $\mathcal{P}^N_M$ are then obtained as follows: the central value for $a_\mu^{\rm HVP,\,LO}$ is the median of the distribution of the results and the uncertainty is obtained within a 68\% confidence level (CL).

Given the size of the errors present in the data sets, now we are only able to obtain meaningful results from approximants with at most 6 fit parameters, which corresponds to the PAs $P_1^1(x)$, $P_1^2(x)$, $P_2^2(x)$, and $P_2^3(x)$, and the D-Logs $D_2^1(x)$, $D_2^2(x)$, $D_3^2(x)$, and $D_3^3(x)$. Higher-order approximants have an excessive number of parameters to fit which leads to gigantic uncertainties and unstable results. As an example of typical fit outcomes, for one of the 1000 data sets, we compare in Fig.~\ref{fig:pasdata} the data and the four PAs obtained after fitting to these data.
We can notice that all approximants fit the data very well and no significant deviation between the approximants is present in the region where data is available. Similar results are found for D-Logs.
Discrepancies in the region where extrapolation is necessary are, however, non-negligible. As discussed in Secs.~\ref{sec:theoryHVP} and~\ref{sec:pastaylor}, the change of variable from $t$ to $x$ masks the fact that there is a significant extrapolation being performed and the integration from $x\approx 0.93$, where the data ends, to $x=1$ still gives an important contribution to $a_\mu^{\rm HVP,\,LO}$.

\begin{figure}[t]
\centering
\includegraphics[width=0.5\textwidth]{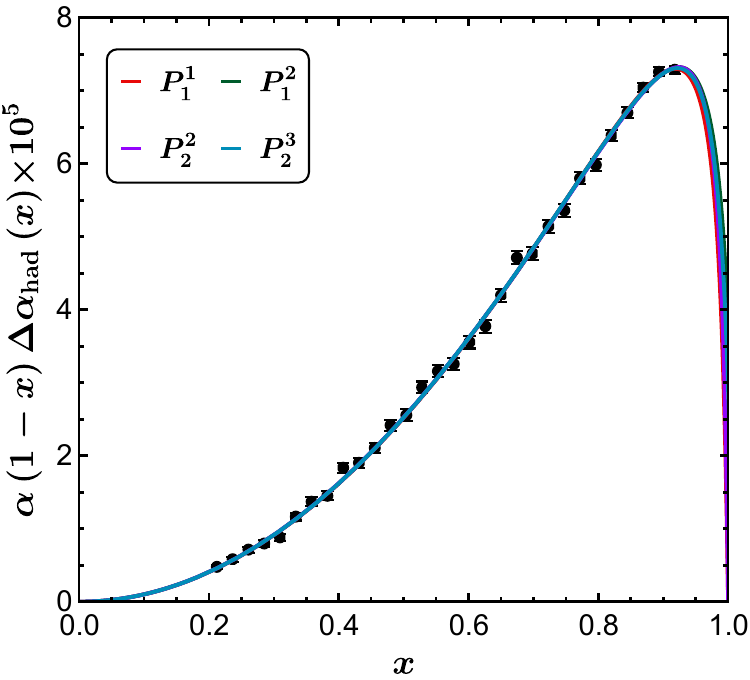}
\caption{Fitted PAs to the integrand $\alpha (1-x) \Delta\alpha_{\mathrm{had}}(x)$ from Eq.~(\ref{eq:amu}). Black points show one toy data set used in the current exercise.}
\label{fig:pasdata}
\end{figure}

Before turning to our final numbers, it is instructive to examine the results from each approximant separately. We give in Tab.~\ref{tab:PAsamu} the values for $a_\mu^{\mathrm{HVP,\,LO}}$ obtained after fitting each PA and D-Log to the 1000 toy data sets, together with the median of the reduced $\chi^2$ (and its 68\% CL). One can notice that the pattern of convergence of Eqs.~(\ref{eq:convergence})~and~(\ref{eq:convergenceDlogs}) is obeyed for the central values obtained from the PAs and D-Logs, as already happened in the idealized case of Sec.~\ref{sec:pasnoerror}. It is also possible to note from Tab.~\ref{tab:PAsamu} that the uncertainty of $a_\mu^{\mathrm{HVP,\,LO}}$ from $P_1^2$ is much larger than the other PA's predictions. This can be understood by the fact that this approximant is not a Stieltjes function, as discussed in Sec.~\ref{sec:theorypas} --- results from $P_1^2$ can, therefore, be safely discarded when confronted with the others. Finally, we observe that the uncertainty of $a_\mu^{\rm HVP,\,LO}$ increases from $P_2^2$ to $P_2^3$ as well as from $D^2_3$ to $D^3_3$ and we could not obtain meaningful fits for $P^3_3$ and $D^3_4$. These are indications that we are at the limit of what can be done with the toy data sets given the size of the errors expected by the MUonE experiment.

The results from Tab.~\ref{tab:PAsamu} show a somewhat large uncertainty. This uncertainty stems mostly from the extrapolation, i.e. deviations of different approximants outside the data region, corresponding to $t \in (-\infty, -0.138]\,\mathrm{GeV}^2$. To quantify the error from this extrapolation, we calculated the partial contributions $a_\mu^{\rm HVP,\,LO}(x_{\rm max})$ up to $x_{\rm max}$, defined in Eq.~(\ref{eq:amuxmax}), with $x_{\mathrm{max}} \in [0.990,1]$. It is expected that by restricting the extrapolation to a smaller interval, the final errors will be smaller. For example, with $x_{\rm max}=0.990$ one can cover 99.1\% of the value of $a_\mu^{\rm HVP,\,LO}$, which may be an acceptable trade-off if the uncertainties are significantly reduced. The remaining 0.9\% would have to be obtained externally by matching to perturbative QCD or lattice data if available at similar precision, for example.

\begin{table}[!t]
\begin{center}
\caption{Results for $a_\mu^{\mathrm{HVP,\,LO}}$ from the PAs and D-Logs used as fitting functions to toy data sets together with the final values for $\chi^2/{n_{\mathrm{dof}}}$ of the respective approximants. Final results for both methods are also presented.}
\begin{tabular}{ccc|ccc}
\toprule
& $a_\mu^{\mathrm{HVP,\,LO}} \times 10^{11}$ & $\chi^2/{n_{\mathrm{dof}}}$ & & $a_\mu^{\mathrm{HVP,\,LO}} \times 10^{11}$ & $\chi^2/{n_{\mathrm{dof}}}$ \\ \hline
$P_1^1$ & $6938 \pm 21$ & $1.01_{-0.25}^{+0.27}$ & $D_2^1$ & $7052^{+66}_{-71}$ & $1.01_{-0.26}^{+0.26}$ \\[0.1cm]
$P_1^2$ & $7042^{+114}_{-104}$ & $1.01_{-0.26}^{+0.28}$ & $D_2^2$ & $6956^{+96}_{-65}$ & $1.05_{-0.27}^{+0.28}$ \\[0.1cm]
$P_2^2$ & $6980^{+46}_{-34}$ & $1.05_{-0.27}^{+0.29}$ & $D_3^2$ & $6999^{+48}_{-39}$ & $1.10_{-0.28}^{+0.29}$ \\[0.1cm]
$P_2^3$ & $6994^{+85}_{-49}$ & $1.11_{-0.31}^{+0.29}$ & $D_3^3$ & $6977^{+72}_{-53}$ & $1.14_{-0.29}^{+0.30}$ \\ \hline
Final result & $6987^{+46}_{-34}$ & --- & Final result & $6988^{+48}_{-39}$ & --- \\ \bottomrule
\label{tab:PAsamu}
\end{tabular}
\end{center}
\end{table}

\begin{figure}[!t]
\centering
\subfloat[]{\label{fig:amupas099}{\includegraphics[width=0.49\textwidth]{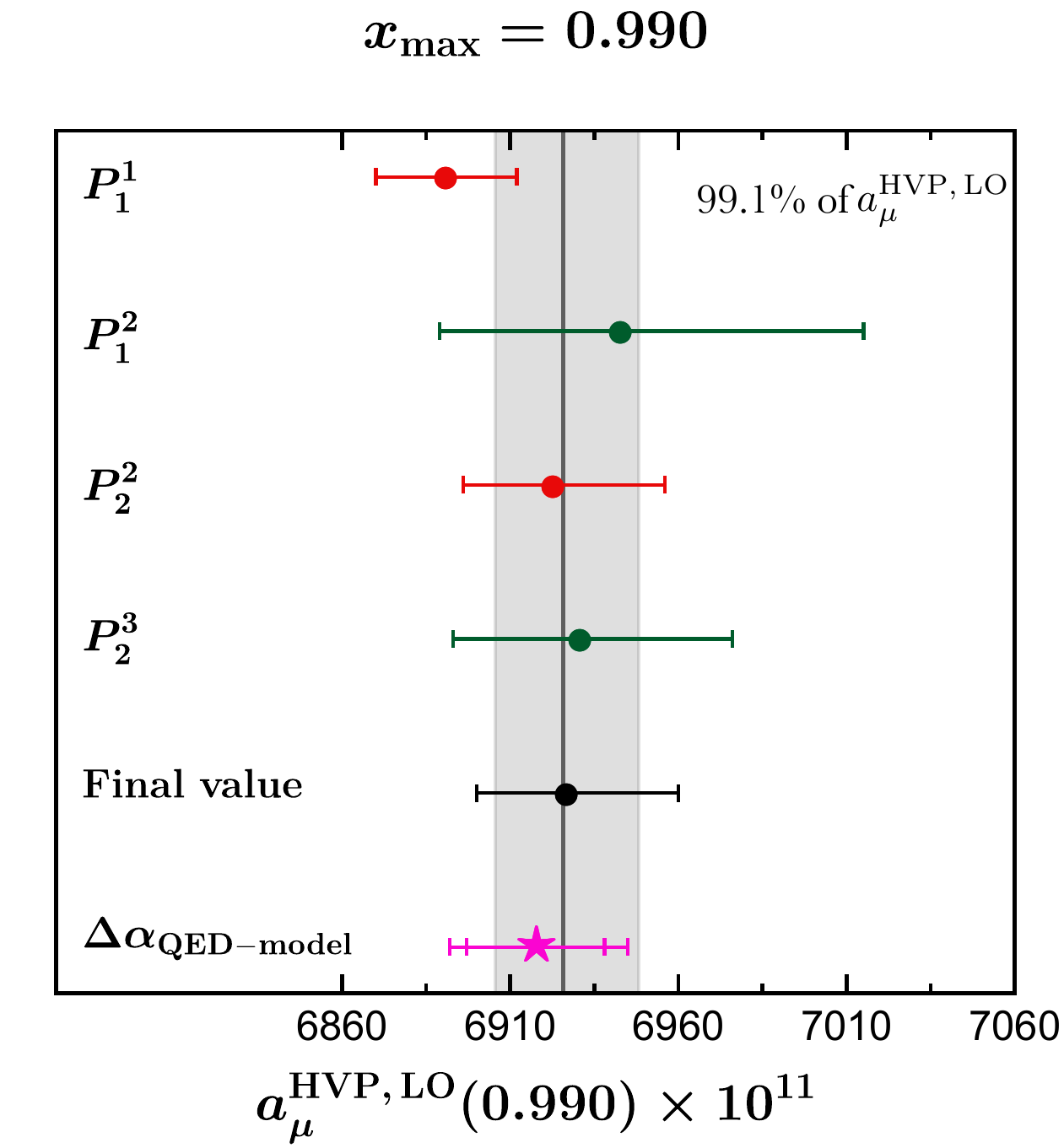}}}\hfill
\subfloat[]{\label{fig:amupas0995}{\includegraphics[width=0.49\textwidth]{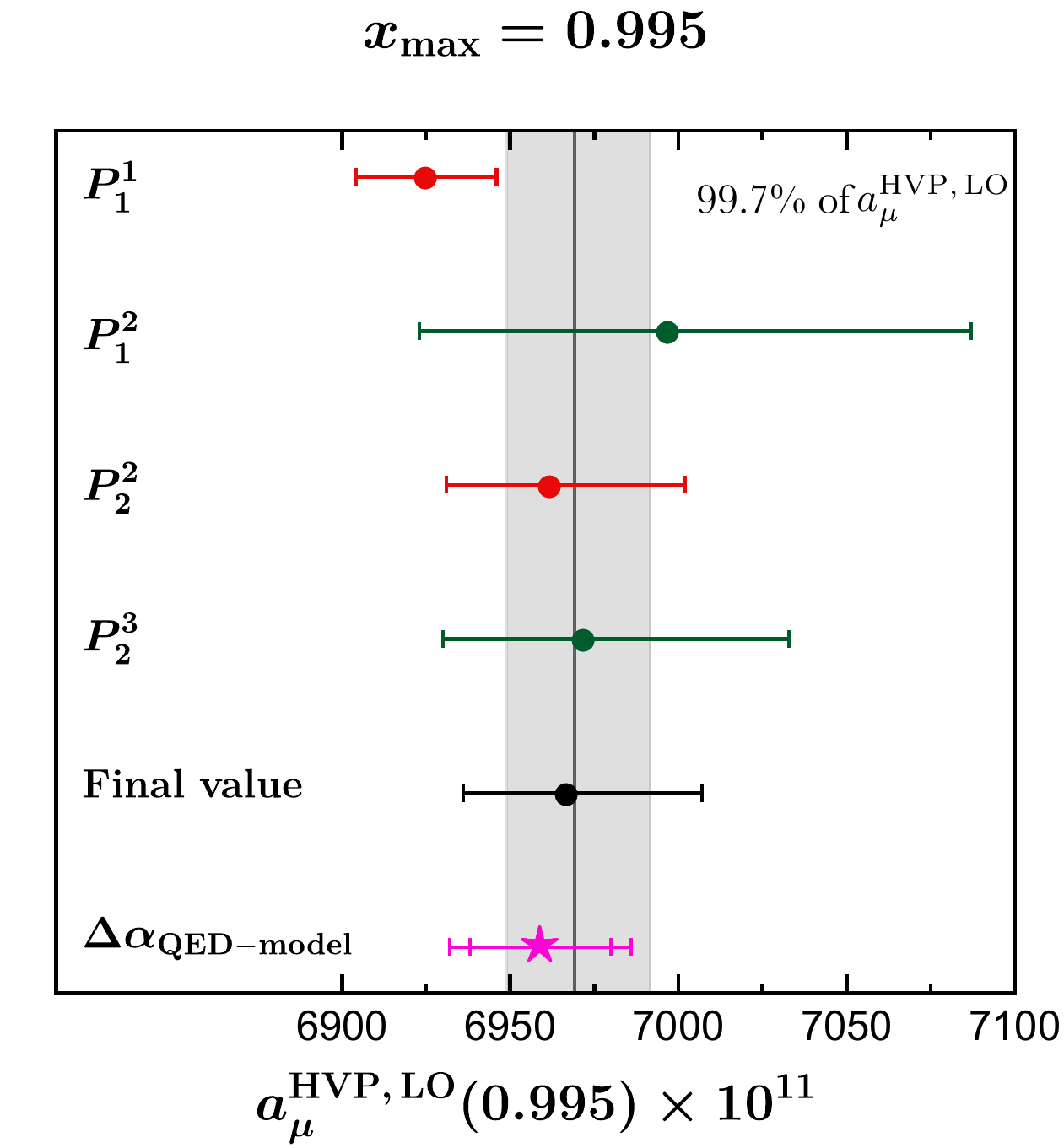}}}\hfill
\subfloat[]{\label{fig:amupas0997}{\includegraphics[width=0.49\textwidth]{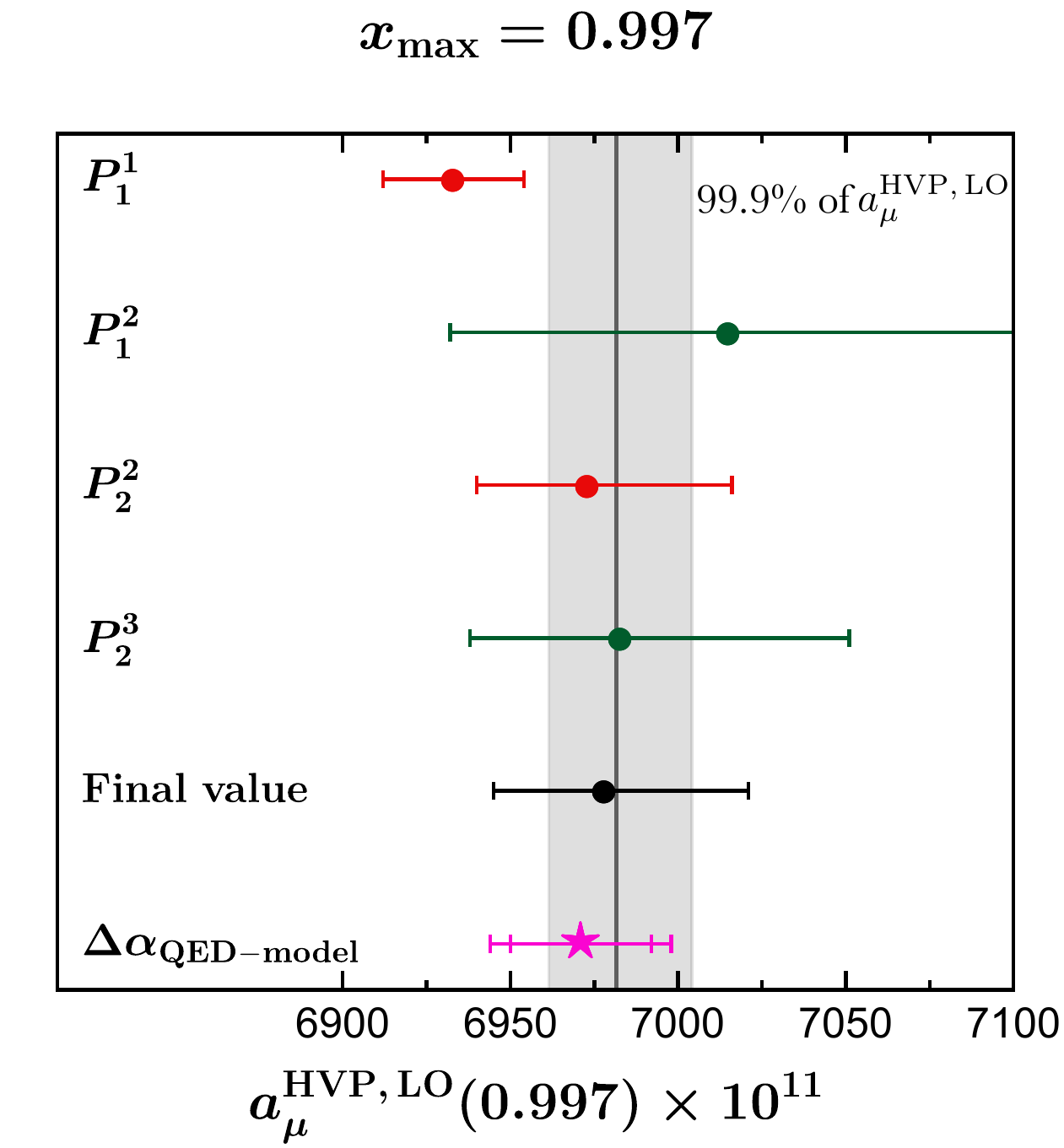}}}\hfill
\subfloat[]{\label{fig:amupas1}{\includegraphics[width=0.49\textwidth]{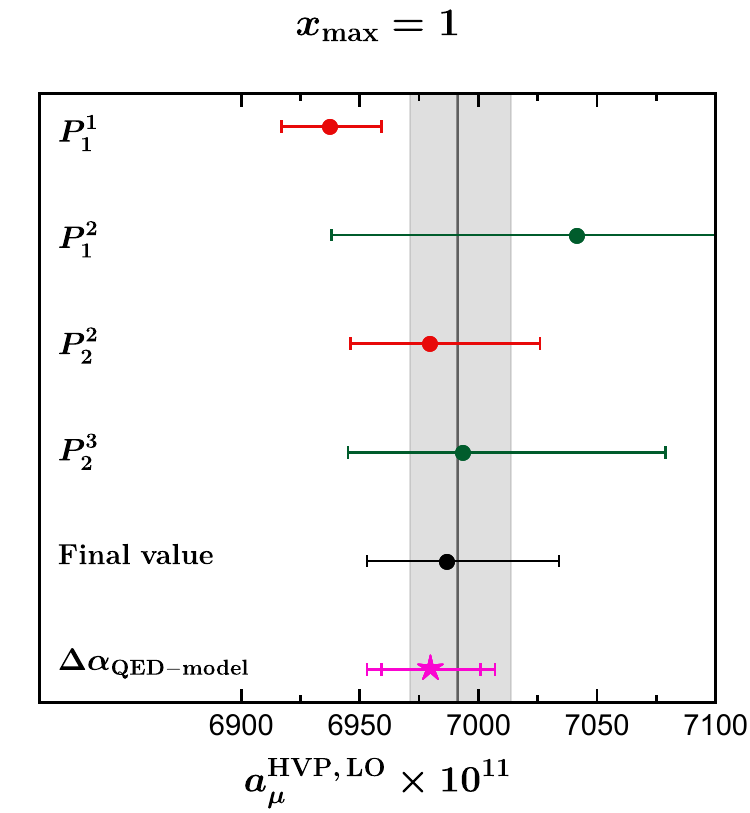}}}
\caption{$a_\mu^{\mathrm{HVP,\,LO}}(x_{\mathrm{max}})$ from PAs fitted to the toy data sets for four different values of $x_{\mathrm{max}}$: (a) $x_{\mathrm{max}}=0.990$ $(t_{\mathrm{max}} = -1.1\,\mathrm{GeV}^2)$, (b) $x_{\mathrm{max}}=0.995$ $(t_{\mathrm{max}} = -2.2\,\mathrm{GeV}^2)$, (c) $x_{\mathrm{max}}=0.997$ $(t_{\mathrm{max}} = -3.7\,\mathrm{GeV}^2)$ and (d) $x_{\mathrm{max}}=1$. The PAs $P_N^N$ are shown in red while $P_N^{N+1}$ appear in green. Final results obtained from the approximants appear as a black dot and results from the QED-inspired model of Eq.~(\ref{eq:deltaalphaqed}) as a pink star. The inner error bar in the QED model result represents the statistical uncertainty. The gray band gives $a_\mu^{\mathrm{HVP,\,LO}}(x_{\mathrm{max}})$ with exact extrapolation using the model of Eq.~(\ref{eq:model}).}
\label{fig:amupas}
\end{figure}

\begin{figure}[!t]
\centering
\subfloat[]{\label{fig:amudlogs099}{\includegraphics[width=0.48\textwidth]{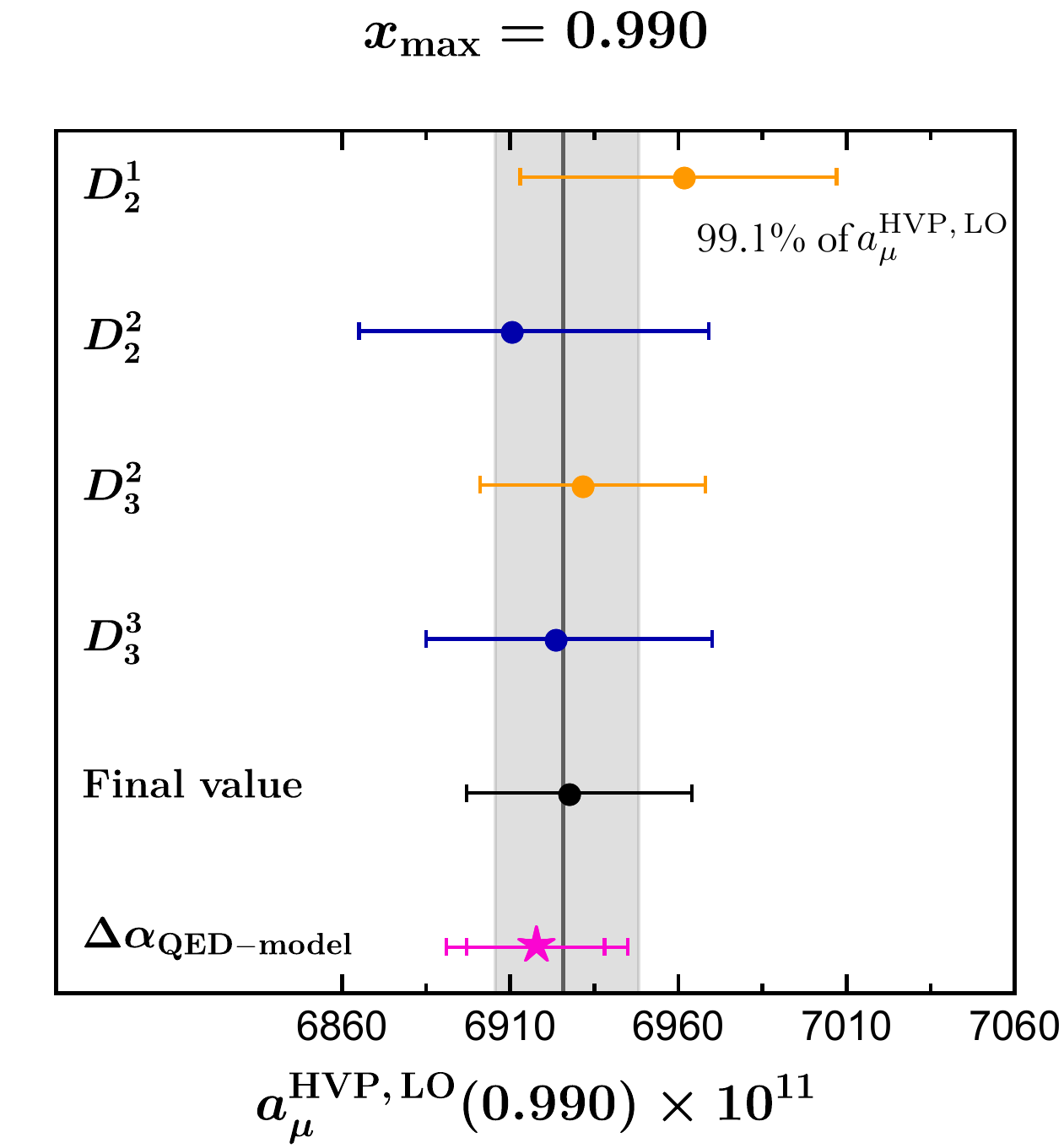}}}\hfill
\subfloat[]{\label{fig:amudlogs0995}{\includegraphics[width=0.48\textwidth]{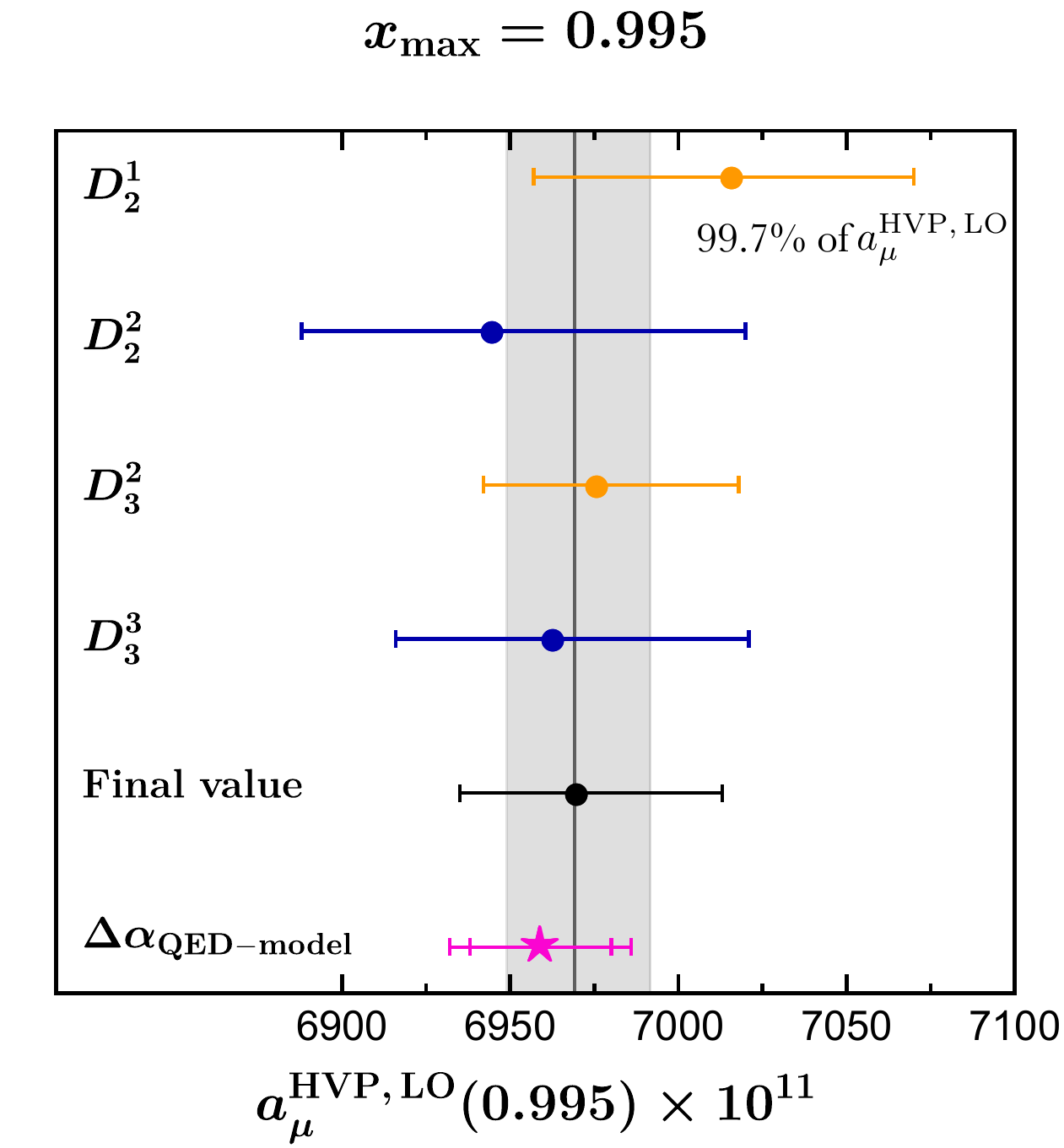}}}\hfill
\subfloat[]{\label{fig:amudlogs0997}{\includegraphics[width=0.48\textwidth]{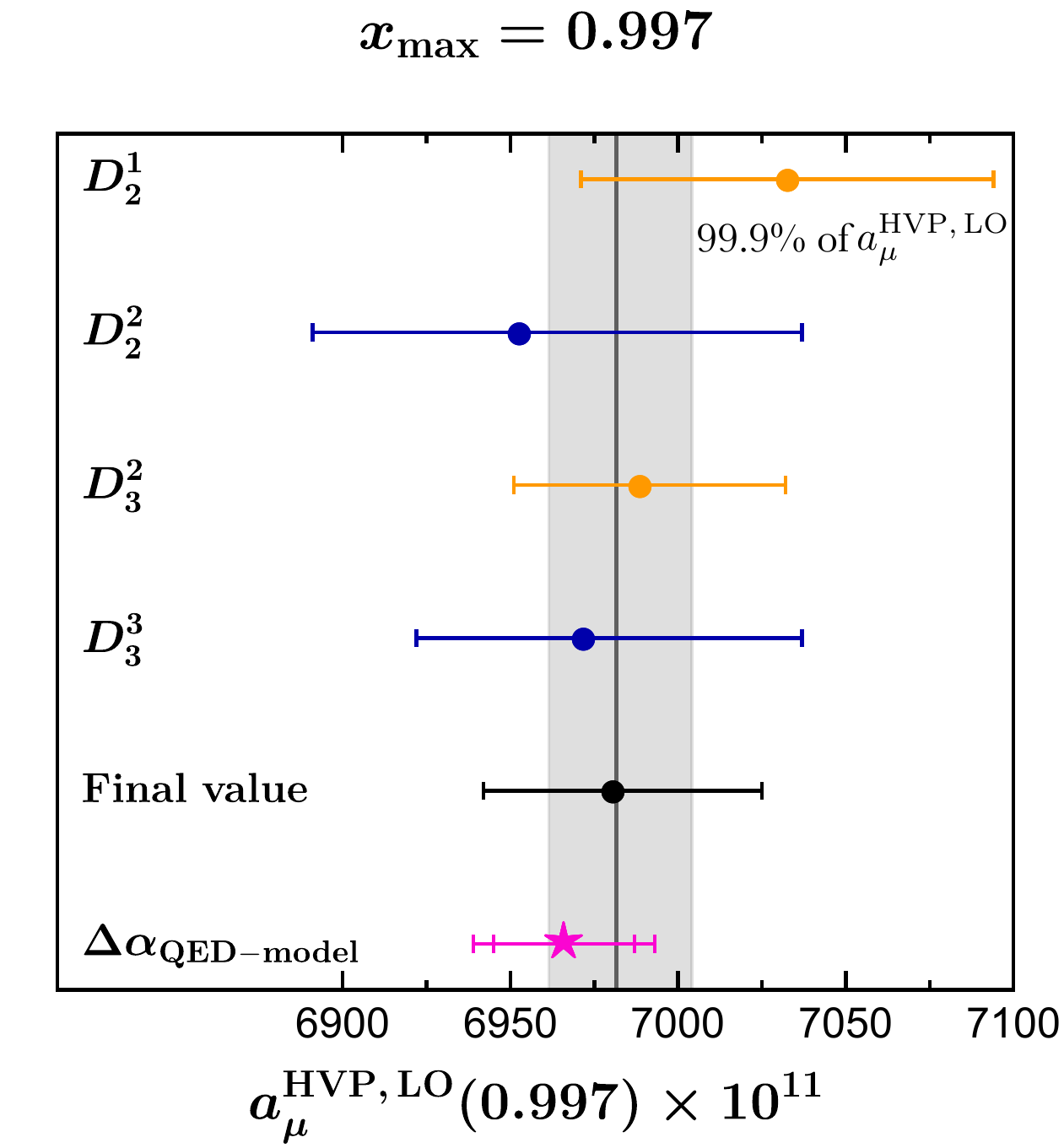}}}\hfill
\subfloat[]{\label{fig:amudlogs1}{\includegraphics[width=0.48\textwidth]{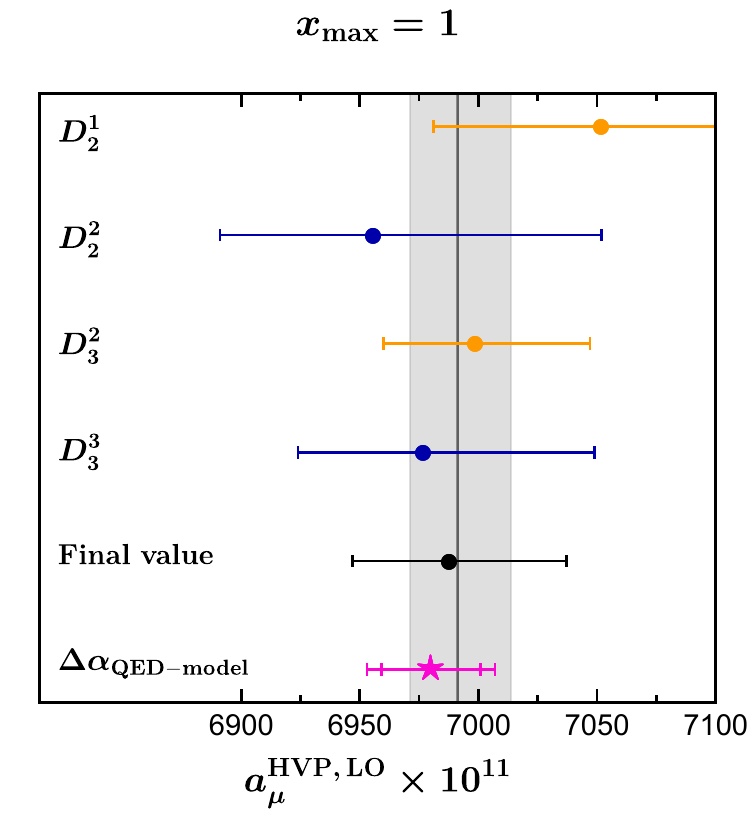}}}
\caption{$a_\mu^{\mathrm{HVP,\,LO}}(x_{\mathrm{max}})$ from D-Logs fitted to the toy data sets for four different values of $x_{\mathrm{max}}$: (a) $x_{\mathrm{max}}=0.990$ $(t_{\mathrm{max}} = -1.1\,\mathrm{GeV}^2)$, (b) $x_{\mathrm{max}}=0.995$ $(t_{\mathrm{max}} = -2.2\,\mathrm{GeV}^2)$, (c) $x_{\mathrm{max}}=0.997$ $(t_{\mathrm{max}} = -3.7\,\mathrm{GeV}^2)$ and (d) $x_{\mathrm{max}}=1$. The D-Logs $D_N^N$ are shown in blue while $D_{N+1}^N$ appear in yellow. Final results obtained from the approximants appear as a black dot and results from the QED-inspired model of Eq.~(\ref{eq:deltaalphaqed}) as a pink star. The inner error bar in the QED model result represents the statistical uncertainty. The gray band gives $a_\mu^{\mathrm{HVP,\,LO}}(x_{\mathrm{max}})$ with exact extrapolation using the model of Eq.~(\ref{eq:model}).}
\label{fig:amudlogs}
\end{figure}

The $a_\mu^{\mathrm{HVP,\,LO}}(x_{\mathrm{max}})$ values estimated for $x_{\mathrm{max}}= \{0.990, 0.995, 0.997,1\}$ are illustrated for the PAs in Fig.~\ref{fig:amupas} and for the D-Logs in Fig.~\ref{fig:amudlogs}.
The pattern of convergence is evident for all $x_{\mathrm{max}}$ in Figs.~\ref{fig:amupas} and~\ref{fig:amudlogs}, where the gray band is the expected model value of $a_\mu^{\mathrm{HVP,\,LO}}(x_{\mathrm{max}})$ given by Eq.~(\ref{eq:model}). The red points in Fig.~\ref{fig:amupas} represent the $P_N^N$ sequence and the green ones $P_N^{N+1}$. In Fig.~\ref{fig:amudlogs}, the blue points show the $D_N^N$ sequence while results for $D_{N+1}^N$ appear in yellow. One can observe in the results of these figures that the uncertainty steadily increases with $x_{\rm max}$, which reflects the dispersion due to extrapolation outside the data region.

We can then obtain a final value for $a_\mu^{\rm HVP,\,LO}(x_{\rm max})$. Since with both PA and D-Logs the approximants from the two sequences are expected to bound the true value, it is natural to use the average of the highest-order approximants, in this case $P_2^2$ and $P_2^3$, on the one hand, and $D_3^2$ and $D^3_3$ on the other, as the central value for the final estimate. We will consider the final statistical uncertainty to be the smallest between each pair of approximants. This can be considered conservative since we do not reduce the final error, as would be done in a weighted average, due to the expected strong correlations between the two results. It is important to mention that
we book the error stemming from the extrapolation ---which is the dominant source of error--- as part of the statistical uncertainty since, ultimately, this error is rooted in the statistical fluctuations of the data. Finally, for the systematic uncertainty we use half the interval spanned by the central values from the two highest-order approximants. Our final estimates for PAs and D-Logs for the different values of $x_{\mathrm{max}}$ are collected in Tab.~\ref{tab:amuuptoxmax} ---second and third columns, respectively--- where ``stat'' and ``sys'' refer to the statistical (in the sense explained above) and systematic uncertainties. These final estimates are also illustrated in Figs.~\ref{fig:amupas}~and~\ref{fig:amudlogs} by the black dots.

As one can notice, our final results are in excellent agreement with the expectation from the model we used (fifth column of Tab.~\ref{tab:amuuptoxmax}), with the central values for $a_\mu^{\rm HVP, LO}$ off by at most 0.06\% in the case of PAs and $0.05 \% $ in the case of D-Logs.
Uncertainties are dominated by statistics and the extrapolation (which we book as ``stat''). A reduction in the uncertainty of about 25\% in both methods is achieved by computing the integral up to $x=0.990$, which covers 99.1\% of $a_\mu^{\rm HVP,\, LO}$. This is very likely an acceptable compromise since estimating the remaining 0.9\% from $e^+e^-$ data or perturbative QCD would not increase the uncertainties in any significant way. Finally, we observe that the systematic uncertainties, which are small, do not change significantly with $x_{\rm max}$.

\begin{table}[!t]
\begin{center}
\caption{Final results for $a_\mu^{\mathrm{HVP,\,LO}}(\pm\mathrm{stat})(\pm\mathrm{sys})$ in units of $10^{11}$ for PAs, D-Logs and the QED-model. ``stat'' and ``sys'' refer to the statistical and systematic uncertainties, respectively.}
\begin{tabular}{ccccc}
\toprule
$x_{\mathrm{max}}$ & $a_{\mu,\,{\mathrm{PAs}}}^{\mathrm{HVP,\,LO}}$ & $a_{\mu,\,{\mathrm{Dlogs}}}^{\mathrm{HVP,\,LO}}$ & $a_{\mu,\,{\mathrm{QED-model}}}^{\mathrm{HVP,\,LO}}$ & $a_{\mu,\,{\mathrm{data-sets}}}^{\mathrm{HVP,\,LO}}$ \\ \midrule
$0.990$ & $6927 \left(^{+33}_{-27}\right)(\pm 4)$ & $6928 \left(^{+36}_{-31}\right)(\pm 4)$ & $6918 \left(^{+21}_{-20}\right)(\pm 4)$ & $6926 \left(^{+22}_{-20}\right)$ \\[0.1cm]
$0.995$ & $6967 \left(^{+40}_{-31}\right)(\pm 5)$ & $6970 \left(^{+42}_{-34}\right) (\pm 7)$ & $6959 (\pm 21) (\pm 17)$ & $6969 \left(^{+22}_{-20}\right)$ \\[0.1cm]
$0.997$ & $6978 \left(^{+43}_{-33}\right)(\pm 5)$ & $6981 \left(^{+43}_{-38}\right)(\pm 9)$ & $6971 (\pm 21) (\pm 17)$ & $6982 \left(^{+22}_{-20}\right)$ \\[0.1cm]
$1.000$ & $6987 \left(^{+46}_{-34}\right)(\pm 7)$ & $6988 \left(^{+48}_{-39}\right)(\pm 11)$ & $6980 (\pm 21) (\pm 17)$ & $6991 \left(^{+22}_{-20}\right)$ \\ \bottomrule
\label{tab:amuuptoxmax}
\end{tabular}
\end{center}
\end{table}

For comparison, we also employ the QED-inspired model of Eq.~(\ref{eq:deltaalphaqed}) to perform the fits to the toy data sets. For every data set a value for the parameters $K$ and $M$ is determined as well as a prediction for $a_\mu^{\mathrm{HVP,\,LO}}(x_{\mathrm{max}})$, obtained by using the fitting function to compute the integral of Eq.~(\ref{eq:amuxmax}).
We again quote the median as the central value with an uncertainty obtained within its 68\% CL.
The predictions for the different $x_{\mathrm{max}}$ are indicated in Tab.~\ref{tab:amuuptoxmax}, where the systematic error is determined as $0.24\%$ of the central value, as explained in Sec.~\ref{sec:pasnoerror}.
These values can be seen in Figs.~\ref{fig:amupas} and~\ref{fig:amudlogs} represented by the pink star with the inner error band representing solely the statistical uncertainty and the larger error band representing the total uncertainty, from the sum in quadrature of the two error sources in Tab.~\ref{tab:amuuptoxmax}. The QED-inspired model, which has only two free parameters, clearly outperforms $P_1^1$, which also has only two free parameters. This is certainly a result of the additional structure of the model, that contains, for example, a logarithmic cut. On the other hand, the result from the model underestimates the true value, which reflects a systematic uncertainty associated with the model dependency, something already observed in Sec.~\ref{sec:pasnoerror}. The model dependency can be inferred from the fitted values of $K$ and $M$ which read $6871^{+43}_{-38} \, \mathrm{GeV}^{-2}$ and $0.060^{+0.004}_{-0.003}\, \mathrm{GeV}^2$, respectively. In particular, the value of $M$ is incompatible with $m_\pi^2$, as would be expected from the model in Eq.~(\ref{eq:model}). In comparison with the final result from the PAs and D-Logs, the model displays a smaller uncertainty, which stems from the fact that it has only two free parameters, but is further away from the true value. In this respect, the use of PAs and D-Logs helps reducing the final systematic uncertainty, being, in addition, completely model independent.

\section{Conclusion}\label{sec:conclusion}

In this work, we described the use of Pad\'e approximants and D-Log Pad\'e approximants to fit and extrapolate toy data reflecting the expected results of the future MUonE experiment. This is a model-independent strategy to extract $a_\mu^{\mathrm{HVP,\,LO}}$ that relies on general knowledge about the fundamental properties of the hadronic contribution to the running of the fine-structure constant, which is a Stieltjes function in the variable $t$~\cite{Aubin:2012me}. These type of functions are ruled by a determinant condition, given in Eq.~(\ref{eq:determinant}), which generates constraints for their Taylor series coefficients. This is a key factor, since the convergence of rational approximants built from the Taylor series of Stieltjes functions is guaranteed by theorems~\cite{Baker1975essentials,Baker1996pade} in a very specific pattern, where the PAs and D-Logs belonging to the diagonal sequence approach $\Delta\alpha_{\mathrm{had}}(t)$ from below while the PAs of the sequence $P_N^{N+1}(t)$ and the D-Logs of the sequence $D_{N+1}^N(t)$ do so from above, as indicated in Eqs.~(\ref{eq:convergence})~and~(\ref{eq:convergenceDlogs}).

In our case, the approximants are used as fitting functions that would be employed to fit the future MUonE data. Our fitting functions were constructed as follows. First, the approximants were formally built with the canonical procedure, i.e. by matching to the Taylor series of $\Delta\alpha_{\mathrm{had}}(t)$ around $t=0$, with generic Taylor coefficients that would be fixed by the data. Then, the function in the variable $x$, which is more suited for fits in the MUonE framework, was obtained by the change of variables of Eq.~(\ref{eq:ttox}). These new functions in $x$ are then used to fit the toy data sets in order to determine $a_\mu^{\mathrm{HVP, \,LO}}$, by employing the results from the approximant in the integral of Eq.~(\ref{eq:amu}) and, as a by product, to estimate the Taylor coefficients of $\Delta\alpha_{\mathrm{had}}(t)$. The toy data sets used in this work were generated from the model for the Euclidean correlator proposed by Greynat and de Rafael in Ref.~\cite{Greynat:2022geu}, briefly motivated in Sec.~\ref{sec:model}. We believe this simple model to be sufficient for our purposes since it captures the main features of $\Delta\alpha_{\mathrm{had}}(t)$ and is a Stieltjes function.

First we showed, as proof of concept, how sequences of approximants bound the true value of $a_\mu^{\rm HVP,\,LO}$ in idealized scenarios with no fluctuations in the data sets. We then turned to tests of our method in the realistic case, where the toy data sets are generated from the model with fluctuations that reflect the expected uncertainties to be obtained from the MUonE experiment. We produced 1000 data sets and fitted them with the PAs $P_1^1$, $P_1^2$, $P_2^2$, and $P_2^3$, and the D-Logs $D_2^1$, $D_2^2$, $D_3^2$, and $D_3^3$. The constraints imposed on the Taylor series coefficients, derived from the determinant condition of Eq.~(\ref{eq:determinant}) and given in Eq.~(\ref{eq:coeffhierarchy}), were imposed on the fits. We analyzed then the estimate of $a_\mu^{\mathrm{HVP, \,LO}}(x_{\rm max})$ for each approximant up to $x_{\mathrm{max}} = $ 0.990, 0.995, 0.997 and 1. Taking advantage of the expected convergence pattern, i.e. the fact that different sequences bound the true value as in Eqs.~(\ref{eq:convergence}) and~(\ref{eq:convergenceDlogs}), we obtained the final estimate of $a_\mu^{\mathrm{HVP, \,LO}}(x_{\mathrm{max}})$. Our final results are obtained from the average between the two highest-order approximants of each sequence, in concrete, the PAs $P_2^2$ and $P_2^3$ and the D-Logs $D_3^2$ and $D_3^3$. The systematic uncertainty of our final prediction can be estimated and it was calculated as half of the difference between these two approximants while the statistical error was taken as the smallest error among the highest-order approximants of each sequence.

For all the values of $x_{\mathrm{max}}$ employed, our final estimates are fully compatible with what was expected from the underlying model, as one can see in Figs.~\ref{fig:amupas}~and~\ref{fig:amudlogs} for the PAs and D-Logs respectively. Our final central values differ from the expected ones by less than 0.06\% in the PAs case and 0.05\% for D-Logs. The final uncertainty is, in all cases, dominated by the uncertainty stemming from the extrapolation. We observe, however, that when the region in which we extrapolate the results is expanded, the dispersion between different fits grows, leading to larger final uncertainties. This can certainly be expected since the extrapolation that is performed is far from trivial and relies on data in a very limited region of the variable $t$. In this respect, an extrapolation in a somewhat limited interval, with $x_{\rm max}=0.990$ for example, which covers $99.1\%$ of the $a_\mu^{\rm HVP,\,LO}$ integrand, significantly reduces the final uncertainty. Of course, in this case one needs to resort to external information to complete the determination of $a_\mu^{\rm HVP,\,LO}$.

We also compared our final prediction with the ones obtained from the QED-inspired model of Eq.~(\ref{eq:deltaalphaqed}) used in preliminary studies of the MUonE proposal~\cite{Abbiendi:2677471}. This model leads to smaller uncertainties than those obtained from our procedure, in part because it has only two parameters, but its central value is further away from the true value, as seen in Figs.~\ref{fig:amupas}~and~\ref{fig:amudlogs}. This larger systematic uncertainty reflects a model dependency that can hardly be avoided with functions of this type.

In summary, in this work we showed that the systematic use of PAs and D-Logs as model-independent fitting functions to the future MUonE data can provide a powerful framework for the extraction of $a_\mu^{\rm HVP,\,LO}$. Our explorations show that this method is superior to the use of a single, fixed, fitting function, which may carry a model dependence and an associated systematic uncertainty that would be difficult to estimate on the basis of real experimental data. The nicest feature of the method is the fact that we expect different sequences to bound the true value, which renders the average of results from these two sequences superior to the estimate arising from a single approximant.
Therefore, the PAs and D-Logs provide the basis for a model-independent and systematic method, relying only on the analytical structure of the two-point correlator underlying $a_\mu^{{\mathrm{HVP,\,LO}}}$, that is able to yield a result with a competitive, although conservative, uncertainty.

\section*{Acknowledgements}
The authors thank Carlo Carloni Calame, Giovanni Abbiendi and Graziano Venanzoni for providing us with the data errors expected in the MUonE experiment. We thank J\'er\^ome Charles for comments on a previous version of this manuscript. PM thanks IFSC, at University of Sao Paulo, and CYL thanks IFAE, at Universitat Aut\`onoma de Barcelona, for hospitality.  The work of CYL was financed by the S\~ao Paulo Research Foundation (FAPESP) grant No.~2020/15532-1 and 2022/02328-2. DB’s work was supported by FAPESP grant No.~2021/06756-6 and by CNPq grant No.~308979/2021-4. The work of PM has been supported by the Ministerio de Ciencia e Innovación under grant PID2020-112965GB-I00 and by the Secretaria d’Universitats i Recerca del Departament d’Empresa i Coneixement de la Generalitat de Catalunya under grant 2021 SGR 00649. IFAE is partially funded by the CERCA program of the Generalitat de Catalunya.

\appendix

\section{Fitting functions constructed using PAs and D-Logs}\label{app:pasfunctions}

In this appendix we give the explicit expressions of the approximants used in this work as a function of $x$ and in terms of the unknown Taylor coefficients $a_n$ of $\Delta\alpha_{\mathrm{had}}$.

We start with the sequence $P_N^N$, where the first PA is $P_1^1$. As seen in Sec.~\ref{sec:pasnoerror}, the fitting function is
\begin{equation}
P_1^1(x) = - \dfrac{b_1 \, m_\mu^2 \, x^2}{1 - x + b_2 \, m_\mu^2 \, x^2} \,,
\end{equation}
with $b_1 = a_1$ and $b_2 = {a_2}/{a_1}$. The constraints employed are $b_1 < 0$ and $b_2 > 1$. The next approximant in this sequence is $P_2^2$, whose final expression is
\begin{equation}
P_2^2(x) = \dfrac{b_1 \, m_\mu^2 \, x^2 \, (x - 1) + (b_2 - b_1 b_3) \, m_\mu^4 \, x^4}{(1 - x)^2 + b_3 \, m_\mu^2 \, x^2 \, (1-x) + b_4 \, m_\mu^4 \, x^4} \,,
\end{equation}
where the fit parameters are now
\begin{equation}
b_1 = a_1, \qquad \quad b_2 = a_2, \qquad \quad b_3 = \dfrac{a_2 a_3 - a_1 a_4}{a_2^2 - a_1 a_3}, \qquad \quad b_4 = \dfrac{a_3^2 - a_2 a_4}{a_2^2 - a_1 a_3} \,.
\end{equation}
From the structure of $\Delta\alpha_{\mathrm{had}}$ and its series representation in Eq.~(\ref{eq:seriesdeltaalpha}), we know that $b_1 < 0$ and $b_2 < b_1$. Analyzing the Stieltjes determinants of Eq.~(\ref{eq:determinant}), we get the additional relations: $b_3 >0$ and $b_4 > 0$. Since the other approximants of the sequence $P_N^N$ were not applied to the realistic data sets, we will refrain from showing their expressions here.

The $P_N^{N+1}$ sequence starts with $P_1^2$, which reads, as a function of $x$
\begin{equation}
P_1^2(x) = - \dfrac{b_1 \, m_\mu^2 \, x^2 \, (1 - x) + (b_1 b_3 - b_2) \, m_\mu^4 \, x^4}{(1 - x)^2 + b_3 \, m_\mu^2 \, x^2 \, (1 - x)} \,.
\end{equation}
The $b_n$ parameters in this case, together with their limits, are
\begin{equation}
b_1 = a_1 < 0, \qquad \qquad b_2 = a_2 < b_1, \qquad \qquad b_3 = \dfrac{a_3}{a_2} > 1.
\end{equation}
The last approximant used is $P_2^3$, which is given by
\begin{equation}
P_2^3(x) = - \dfrac{b_1 \, m_\mu^2 \, x^2 \, (1 - x)^2 - (b_2 + b_1 b_4) \, m_\mu^4 \, x^4 \, (1 - x) + (b_3 + b_1 b_5 + b_2 b_4) \, m_\mu^6 \, x^6}{(1 - x)^3 - b_4 \, m_\mu^2 \, x^2 \, (1 - x)^2 + b_5 \, m_\mu^4 \, x^4 \, (1 - x)} \,,
\end{equation}
where the parameters are
\begin{equation}
b_1 = a_1, \qquad b_2 = a_2, \qquad b_3 = a_3, \qquad b_4 = \dfrac{a_2 a_5 - a_3 a_4}{a_3^2 - a_2 a_4}, \qquad b_5 = \dfrac{a_4^2 - a_3 a_5}{a_3^2 - a_2 a_4}.
\end{equation}
In addition, the constraints employed in the fits are: $b_3 < b_2 < b_1 < 0$, $b_4 < 0$ and $b_5 > 0$. It is important to stress that all the constraints showed in this Appendix are model independent, since they follow from the fact that $\Delta\alpha_{\mathrm{had}}(t)$ is a Stieltjes function.

\begin{table}[!t]
\begin{center}
\caption{D-Log fitting functions as a function of $t$ or $x$.}
\begin{tabular}{cll}
\toprule
& $D^N_M(t)$ & $D^N_M(x)$\\
\midrule
$D_2^1$ & $\frac{-f_0\,t}{(r_1-t)^{\gamma_1}}$ &$\frac{f_0\,m_{\mu}^2\,x^2 (1-x)^{-1+\gamma_1}}{(r_1-r_1x+m_{\mu}^2x^2)^{\gamma_1}}$ \\[0.1cm]
$D_2^2$ & $\frac{-f_0\,t\,e^{\beta t}}{(r_1-t)^{\gamma_1}}$ &$\frac{f_0\,m_{\mu}^2\,x^2 (1-x)^{-1+\gamma_1}}{(r_1-r_1x+m_{\mu}^2x^2)^{\gamma_1}}e^{\beta\frac{m_{\mu}^2x^2}{(x-1)}}$ \\[0.1cm]
$D_3^2$ & $\frac{-f_0\,t}{(r_1-t)^{\gamma_1}(r_2-t)^{\gamma_2}}$ &$\frac{f_0\,m_{\mu}^2\,x^2 (1-x)^{-1+\gamma_1+\gamma_2}}{(r_1-r_1x+m_{\mu}^2x^2)^{\gamma_1}(r_2-r_2x+m_{\mu}^2x^2)^{\gamma_2}}$ \\[0.1cm]
$D_3^3$ & $\frac{-f_0\,t\,e^{\beta t}}{(r_1-t)^{\gamma_1}(r_2-t)^{\gamma_2}}$ &$\frac{f_0\,m_{\mu}^2\,x^2 (1-x)^{-1+\gamma_1+\gamma_2}}{(r_1-r_1x+m_{\mu}^2x^2)^{\gamma_1}(r_2-r_2x+m_{\mu}^2x^2)^{\gamma_2}}e^{\beta \frac{m_{\mu}^2x^2}{(x-1)}}$ \\ \bottomrule
\label{tab:dlogtypefunction}
\end{tabular}
\end{center}
\end{table}

The D-Logs are constructed from the Taylor series as was described in Sec.~\ref{sec:pasnoerror} and reparametrized in terms of the $x$ variable. All the functions employed for fitting purposes are detailed in Tab.~\ref{tab:dlogtypefunction}. Parameters $r_n$ refer to the branch point in the $t$-variable, $\gamma_n$, refers to the multiplicity for the corresponding cut, $\beta$ is an exponential factor that appears only in the diagonal sequence and $f_0$ is a normalization factor.

\addcontentsline{toc}{section}{References}
\bibliographystyle{jhep}
\bibliography{References}

\end{document}